\documentclass[showpacs,pra,floatfix,nofootinbib,reprint]{revtex4-1}
\usepackage[paper=letterpaper]{geometry} 
\usepackage{amsmath}
\usepackage{amssymb}
\usepackage{amsfonts}
\usepackage{delarray}
\usepackage{graphicx}
\usepackage{setspace}
\usepackage{fullpage}
\usepackage{float}
\usepackage{subfigure}
\usepackage{longtable}
\newcommand{\bra}[1]{\langle #1|}
\newcommand{\ket}[1]{|#1\rangle}
\newcommand{\braket}[2]{\langle #1|#2\rangle}
\newcommand{\ketbra}[2]{| #1 \rangle \langle #2 |}
\newcommand{\Tr}{\mathrm{Tr}}

\begin{document}

\title{Kinematic Effect of Indistinguishability and Its Application\\
to Open Quantum Systems}

\author{P. W.~Bryant}
\affiliation{Center for Complex Quantum Systems,
             Department of Physics,
             University of Texas at Austin, Austin, Texas 78712}

\pacs{03.65.Ta,03.65.Yz,34.10.+x}
\date{\today}

\begin{abstract}
In quantum mechanics,
useful experiments require multiple measurements performed on the
identically prepared physical objects composing experimental ensembles.
Experimental systems
also suffer from environmental interference, and one should not
assume that all objects in the experimental ensemble
suffer interference identically from a single, uncontrolled environment.
Here we present a framework for treating multiple quantum environments 
and fluctuations affecting only subsets of the experimental ensemble.
We also discuss a kinematic effect of indistinguishability not applicable
to closed systems.
As an application, we treat inefficient photon scattering as an open system.
We also create a toy model for the environmental interference suffered
by systems undergoing Rabi oscillations, and we find that this
kinematic effect may explain the puzzling Excitation Induced Dephasing
generally measured in experiments.
\end{abstract}

\maketitle

\section{Introduction}
It is because quantum theory makes probabilistic predictions that
all quantitatively useful experiments require repeated measurements
performed on identically prepared physical objects~\cite{dirac_qm_book}.
These identically prepared physical objects
constitute what we will call the experimental ensemble.
And one cannot assume that all ensemble members,
while treated identically by an experimenter,
are also treated identically by an uncontrolled environment.

First we will explain what is meant by the experimental ensemble.
Then to treat multiple quantum environments we will introduce a
framework that one can use in addition to the typical
reduced dynamics of open systems~\cite{petruccione_open_quantum_systems}.
We will also explore a kinematic effect arising from the indistinguishability
of different ensemble members.
As applications, and to isolate the kinematics, we will work
within the framework to model systems with either trivial or simple
closed-system dynamics.
We will treat as an open system the inefficient
scattering of photons from a beam splitter.
Then we will create a toy model for real experimental
systems undergoing Rabi oscillations.
We find that the general yet often puzzling phenomenon called
Excitation Induced Dephasing results from the kinematics of
indistinguishable ensemble members, and consequently we find very good
quantitative agreement with a wide variety of experiments.
For one well-known experiment~\cite{meekhof_rabi_1996}, we find
unprecedented quantitative agreement.

\section{Measurements and the Experimental Ensemble} \label{sec:meas}
The experimental ensemble is particularly easy to identify in
modern experiments, such as~\cite{nagourney_dehmelt_shelved_1986,
bergquist_qjumps_1986,sauter_toschek_qjumps_1986,peik_qjumps_1994,
meekhof_rabi_1996,brune_rabi_1996},
in which one repeatedly prepares, manipulates,
and then measures the state of single physical objects, one at a time.

For example,
in~\cite{meekhof_rabi_1996}, a single beryllium ion is confined in a Paul trap.
It is prepared to be in a motional and hyperfine ground state, and
then a laser drives stimulated Raman transitions for a time duration, $t$,
after which the state of the single ion is measured.
Following that single observation,
the preparation--evolution--measurement process is repeated.
For every value of $t$ at which the experiment is performed, the result
is the average of 4000 observations~\cite{wineland_experimental_1998}.
There is an experimental ensemble for every measured value of $t$, and
in this example every ensemble has size 4000.

We offer these experiments only as clear illustrations of the experimental
ensemble
(though we will address~\cite{meekhof_rabi_1996}
and~\cite{brune_rabi_1996} again in Section~\ref{sec:ap}).
Often multiple objects from the experimental ensemble
are present simultaneously in the laboratory.

To generalize for systems and observables with discrete spectra,
let $\lambda$ label a possible result of a quantum mechanical measurement.
Assume that one performs $N$ measurements, and that 
the result $\lambda$ is found $n_\lambda$ times.
The results of these measurements are counting ratios:
\begin{equation} \label{eq:meas_statement}
\frac{n_\lambda}{N}(t).
\end{equation}
As is clear from experiments, the $t$ in~\eqref{eq:meas_statement}
corresponds to a duration in time
rather than to a time on the laboratory's clock.
Depending on the type of experiment, the ensemble size, $N$, may or may not
be a function of $t$.

The $N$ measurements in~\eqref{eq:meas_statement}
are performed on an experimental ensemble
of $N$ identically prepared physical objects.
Because individual measurements are always assumed to be
independent of each other, the individual members of
the experimental ensemble are assumed not to interact with each other.
Thus the experimental ensemble must not be confused with a many-particle
state having interacting components,
for which useful measurements require an experimental
ensemble of many-particle objects.
A major subject of this study is whether or not the members can be considered
also statistically independent in terms of exchange symmetry or other
correlations.
This will be discussed in detail later.

\section{Kinematic Aspects of Quantum Theory}
\label{sec:rule}
One compares the experimental results in the form of~\eqref{eq:meas_statement} 
to theoretical probabilities.
The theoretical image of any closed, quantum mechanical system is an
operator algebra defined in a linear scalar-product space, $\Phi$.
The vectors $\phi_\alpha$ span $\Phi$, and every linear combination
of the $\phi_\alpha\in\Phi$ can possibly represent the state of the system.
Assume that preparable states are represented by the basis vectors
$\phi_\alpha$.
In practice, one infers from a measurement or from a preparation procedure
the state of an experimental system.
From the $\phi_\alpha$ one constructs the appropriate density operator,
\begin{equation*}
\rho(t)=\sum_{\alpha} w_{\alpha} \ketbra{\phi_\alpha(t)}{\phi_\alpha(t)},
\quad \textrm{with }\sum_{\alpha} w_{\alpha}=1,
\end{equation*}
to represent the inferred state of the closed system.
In this paper, we will only consider systems for which $\alpha$ takes
discrete values.

One characterizes an experimental observable by a prescription for how
it is to be measured,
and in the theory one represents it using a linear operator in $\Phi$.
For simplicity and because it is appropriate for the
experiments described above, we will consider
a projection operator, $\Lambda=\ketbra{\lambda}{\lambda}$,
into the one-dimensional subspace of $\Phi$
that is associated with the observable labeled above by $\lambda$.
If $\lambda$ can be a value of the index $\alpha$, then 
the subspace spanned by $\ketbra{\lambda}{\lambda}$ will be spanned also
by one of the $\ketbra{\phi_{\alpha}}{\phi_{\alpha}}$, such that
the inner product $\braket{\lambda}{\phi_\alpha}\propto\delta_{\alpha\lambda}$.
It is straightforward to generalize to observables that do not have the
$\phi_\alpha$ as eigenvectors.

These are kinematic aspects of quantum theory, and they are
thus independent of any particular quantum system.
The dynamics specific to a particular system enters the theory 
only through a concrete
choice of the space $\Phi$ and of the operator algebra defined in it.

For projection operators such as $\Lambda$,
the expectation value is the probability of finding the discrete value
$\lambda$ in a measurement.
The predictions of quantum theory are calculated in the form of
Born probabilities, $\mathcal{P}_{\Lambda}\big(\rho(t)\big)$, where
\begin{equation*} 
\mathcal{P}_{\Lambda}\big(\rho(t)\big)=\mathrm{Tr}\big(\Lambda\,\rho(t)\big).
\end{equation*}
The explicit comparison between experiment and theory is
\begin{widetext}
\begin{eqnarray} \label{eq:comparison}
\textrm{counting ratio}\equiv
\frac{n_\lambda}{N}(t)
&\stackrel{?}{=}&
\mathcal{P}_{\Lambda}\big(\rho(t)\big)
\equiv\textrm{ Born probability}. \\
\textrm{\textbf{Experiment}} & & \textrm{\textbf{Theory}} \nonumber
\end{eqnarray}
\end{widetext}

When comparing experimental results to theoretical
predictions~\eqref{eq:comparison},
increasing $N$ leads to stronger statistical conclusions regarding an inferred
outcome, but increasing $N$ does not \emph{change} the outcome itself.
In other words, the theoretical predictions on the right hand side
of~\eqref{eq:comparison}
are always independent of the number of members in an ensemble, $N$,
because no theoretical prediction should depend on the number of times 
one performs an experiment to verify it.
A theory's independence from $N$ is a necessary condition for the
identification of the experimental ensemble.

\section{Multiple Quantum Environments for a Single Experiment}
\label{sec:mqe}
Experimental systems cannot be perfectly isolated,
and actual measurements always reveal the experimental signatures of
environmental interference.
In quantum theory there exist well-understood mechanisms to treat
the interference between a quantum system and its
environment~\cite{petruccione_open_quantum_systems}.
The result is often a reduced dynamics, which can be used to 
calculate for an open system an increase in entropy or a loss of coherence.

One has typically not considered the possibility that different
ensemble members might suffer interference from
different environmental systems or from fluctuations within a single
environmental system.
That this possibility is relevant is especially obvious
for the experiments described
above~\cite{nagourney_dehmelt_shelved_1986,
bergquist_qjumps_1986,sauter_toschek_qjumps_1986,peik_qjumps_1994,
meekhof_rabi_1996,brune_rabi_1996},
in which no two ensemble members are ever present simultaneously in the
laboratory.
Of course, because quantum theory describes
physical phenomena independently of how many ensemble members
are present simultaneously, then it must be relevant for all experiments
and any number of simultaneously present members.

By ignoring the possibility that different ensemble members suffer
interference differently, one implicitly assumes either that the uncontrolled
environment treats them all identically or that experimenters
can prepare and control the state of the entire environment.
In practice, this is either unlikely or impossible.

\subsection{Different environments from the beginning}
The simplest possibility for treating multiple quantum environments
is to associate, for the entirety of the experiment,
different ensemble members with different environmental systems.
We will use $m$ to label these different environmental systems, $E_m$.
The density operator representing the state of $E_m$ will be
labeled $\rho_{E_m}$.
The state of the subset of ensemble members suffering interference from the
environmental system $E_m$ will be represented by $\rho_m$.
One must also label the time evolution operator, $U_m(t)$, defined so that
as a function of the duration, $t$, after active preparation,
the time evolution of the composite of system and environment is
\begin{equation*}
U_m(t)\,\rho_m\otimes\rho_{E_m}\, U^\dagger_m(t).
\end{equation*}
For a measurement performed at $t$,
the states of the ensemble members are represented by the density operators
\begin{equation} \label{eq:redtime}
\rho_m(t)=\mathrm{Tr}_{E_m}
\big( U_m(t)\,\rho_m\otimes\rho_{E_m}\, U^\dagger_m(t) \big),
\end{equation}
where $\mathrm{Tr}_{E_m}$ denotes a partial trace over the degrees of freedom
of the environmental system, $E_m$.
This is the typical reduced dynamics for
open systems~\cite{petruccione_open_quantum_systems}, but
we have added the index $m$ to label the dynamics corresponding to the
different environmental systems.

The density operator, $\rho(t)$, to be used in comparison with
experimental data~\eqref{eq:comparison} is then
\begin{equation} \label{eq:type1sum}
\rho(t)=\sum_{m=0}^M\,a_m\,\rho_m(t).
\end{equation}
Here $M$ is the number of different environments deemed necessary
for a model.
So that the left hand side of~\eqref{eq:type1sum} is a valid density operator,
the real numbers $a_m$ in~\eqref{eq:type1sum} must satisfy
the convexity requirements
\begin{equation*}
a_m\geq 0 \quad \textrm{and}\quad \sum_{m=0}^M\,a_m=1.
\end{equation*}

\subsection{Interference events}
\label{sec:ievents}
With~\eqref{eq:type1sum}, any given ensemble member remains for
the entirety of the experiment in the subset corresponding to a given
environmental system.
This will be problematic if there occur fluctuations in 
the environmental systems themselves,
and the fluctuations affect only portions of subsets.
We have assumed that members of the entire ensemble suffer interference
differently,
so we must assume also that members of any subset from~\eqref{eq:type1sum}
suffer differently.
In a general treatment, there will be no fixed subsets of ensemble members.

To treat the possibility that members of \emph{any} subset
suffer environmental interference differently, 
one can assume that different fluctuations occur
during different events throughout an experiment's duration.
Such events could describe something as simple as a temporary
loss of homogeneity of a magnetic field.
Another possibility for an interference event is the actual passive
measurement of an observable in the state of an ensemble member.
In the context of decoherence, Leggett has called this measurement process
``garbling''~\cite{leggett_arrow_qmeasurement}.\footnote{
Dirac has described the act of measurement as causing
``a jump in the state of the dynamical system''~\cite{dirac_qm_book}.
While other interpretations exist, in active measurements
ensemble members are never observed to be in superpositions.
If nature does not distinguish between active and passive measurements,
it is precisely this ``jump in the state'' that requires a new density operator.
}

We imagine the following sequence of interference events.
Recall that, when treating actual data~\eqref{eq:comparison},
and thus when representing the state of ensemble members,
the $t$ represents a duration from active preparation.
All members are therefore initially prepared at $t=0$.
\begin{enumerate}
\item A physicist actively prepares the initial state of all
experimental ensemble members such that they are represented by
a density operator labeled $\rho_0(t)$.
At $t=t^*_1$ there occurs an interference event that
cannot be described by the closed-system dynamics.
If we ignore for now any finite duration of the event and any
other source of environmental interference,
then before the interference event ($0\leq t < t^*_1$)
all ensemble members are represented by $\rho_0(t)$.
\item At $t=t^*_1$ the interference event
affects some subset of the experimental ensemble.
Because the closed-system dynamics cannot describe the interference,
the affected members can no longer be represented by $\rho_0(t)$.
Thus, to represent the state of the affected subset, one requires
a new density operator: $\rho_1(t-t^*_1)$.
The unaffected members will continue to be in a state represented by
$\rho_0(t)$.
\item If there is another interference event occurring at
a duration $t_2^*$, then 
another density operator, $\rho_2(t-t^*_2)$, will be required.
\end{enumerate}

The density operator, $\rho(t)$, to be used in comparison with
experimental data~\eqref{eq:comparison} will thus exhibit a branching behavior
as time progresses:
\begin{equation} \label{eq:cases}
\rho(\tilde{t}) = 
\begin{cases}
\rho_0(t_0) & 0\leq \tilde{t} < t^*_1 \\
a_0\,\rho_0(t_0) + a_1\,\rho_1(t_1) & t^*_1 \leq \tilde{t} < t_2^* \\
a_0\,\rho_0(t_0) + a_1\,\rho_1(t_1) + a_2\,\rho_2(t_2) & t^*_2 \leq \tilde{t}.
\end{cases}
\end{equation}
There is a notational complication here because the time parameter for
experimental results, and thus the $t$ in~\eqref{eq:comparison},
represents a duration from preparation.
We will write $\tilde{t}$ on the left hand side to represent duration
from active preparation of ensemble members.
The branches on the right hand side of~\eqref{eq:cases}
will be functions of the different $t_i=t-t^*_i$, representing
duration from the times $t^*_i$ when the $\rho_i$ became necessary.
If the distinction is unnecessary, we will use $t$.
For closed systems, this notation is not needed because
$t_i\rightarrow t=\tilde{t}$.

Every line in~\eqref{eq:cases} could be written in 
the same form as~\eqref{eq:type1sum}, but here $M$ would be the number
of interference events.
If there is a sequence of such events, say at $t^*_i$,
then the number of density operators on the right hand side
will depend on the time duration, $\tilde{t}$,
and more and more branches will appear as the time duration grows.

Note that nothing new has been introduced to standard quantum mechanics,
and that we have not assumed any new microphysical phenomena.\footnote{
If the interference event is a passive measurement, the state vector 
collapse or wave function collapse~\cite{stamatescu_collapse_2009} mechanism
clearly cannot describe the branching process because it changes the density
operator in~\eqref{eq:cases} in ways that depend on the ensemble size, $N$.
}
In the absence of environmental interference, these
approaches obviously reduce to the theory for closed systems.
In the limit that all members suffer external interference identically,
one has the typical formalism for open systems.
All we have done differently
is apply the standard theory to open systems in such a way
that one can address the possibility that different ensemble members may
suffer environmental interference differently.
One should not read~\eqref{eq:cases}
to mean that the density operator undergoes a particular, dynamical
change.
Rather, this expression means that subsets of ensemble members suffer
differently and therefore require different density operators to
represent their states.

\subsection{The branching process}
To represent the possibility that
between $\tilde{t}$ and $\tilde{t}+\Delta \tilde{t}$
and before active measurement, some subset of the experimental ensemble
suffers interference and afterward
must be described by a new density operator, $\rho_1$,
let us write from~\eqref{eq:cases}
\begin{equation} \label{eq:physguess}
\rho_0(\tilde{t})
\quad\longrightarrow\quad
a_0\,\rho_0(\tilde{t}+\Delta \tilde{t})+a_1\,\rho_1(\tilde{t}+\Delta \tilde{t}).
\end{equation}

Note that~\eqref{eq:physguess} is not mathematically correct.
Let $\rho_i$ be constructed from vectors in the space $\Phi_i$.
In $\Phi_i$ is defined the operator algebra representing the system's dynamics
following one particular type of environmental interference,
which  may not be defined in the space of the closed system itself.
The dynamics of environmental interference is uncontrolled and may be unknown.
In general, $\Phi_i\neq\Phi_{j\neq i}$, and one cannot perform the sum
in~\eqref{eq:physguess}.
Instead,~\eqref{eq:physguess} is 
a phenomenological statement that is required because one can neither perfectly
isolate experimental systems nor ensure that all ensemble members suffer
interference identically.

While~\eqref{eq:physguess} might not make mathematical sense, 
it will have a sensible physical interpretation if
the calculated probabilities are real numbers between 0 and 1.
The corresponding expression for the Born probability is
\begin{eqnarray} \label{eq:branchp}
\mathrm{Tr}\big(\Lambda_0\rho_0(\tilde{t})\big)
& \longrightarrow & 
a_0\,\mathrm{Tr}\big(\Lambda_0\rho_0(\tilde{t}+\Delta \tilde{t})\big) 
\nonumber \\
& &+\, a_1\,\mathrm{Tr}\big(\Lambda_1\rho_1(\tilde{t}+\Delta \tilde{t})\big).
\end{eqnarray}
To calculate Born probabilities one must decide how the operators
$\Lambda_i$ or $\rho_i$ might be extended or limited to other spaces
such that this requirement is fulfilled.
For the remainder of this paper, we will assume that
this can be done.\footnote{
One possibility of interest 
is that $\Lambda_i$ is the zero operator in $\Phi_i$.
In the beam splitter application
we will demonstrate that this corresponds
to inefficiencies in measurement or to dissipative effects.
}
Again, the weights, $a_0$ and $a_1$, in~\eqref{eq:physguess}
and ~\eqref{eq:branchp} must satisfy $a_0+a_1=1$.

We have ignored any dynamics hidden by the arrow in~\eqref{eq:physguess}.
If the timescale associated with the interference event,
$\Delta \tilde{t}$ in~\eqref{eq:physguess},
is very small compared to the dynamical timescales of the experiment,
then one can safely ignore it.
If it is not, then in models one must include interference dynamics along with
the regular system dynamics.

\subsection{Partitioning and indistinguishability}
\label{sec:partind}
Finally, let us consider the weights, $a_i$, in~\eqref{eq:type1sum}
and~\eqref{eq:cases}.
The $a_i$ express the relative sizes of the ensemble's different subsets,
which are in states represented by the different $\rho_i$.
To determine the relative sizes of these subsets,
one must partition an ensemble of arbitrary size, $N$.

To complicate matters, there is a kinematic
requirement that limits one's knowledge of
which members suffer interference from which environments.
When the spatial wave functions of $n$ simultaneously present
ensemble members overlap, any $n$-body wave function must
transform as one of the irreducible representations of the permutation
group of order $(n!)$.
Even if the non-interacting ensemble members are sufficiently
spaced that particle statistics can be ignored, when an external system
cannot distinguish them, interactions still must respect a permutation symmetry.
It is this condition that leads to the well-known phenomenon of
super-radiance~\cite{dicke_superrad_1954}.
This phenomenon has previously been investigated in the context of open
systems~\cite{lidar_decofree_2001} and called collective decoherence,
in which a bath (environment) cannot distinguish between $n$ qubits.
Here we have generalized to allow multiple baths, and we will
attempt to treat the case that $n$ varies and is not known.

Clearly, if one does not know $n$, then one cannot explicitly address
a permutation symmetry when determining the $a_i$.
The physical interpretation of the permutation symmetry is that 
some simultaneously present members of the ensemble
are indistinguishable and cannot be given physically meaningful labels.

We conclude with a set of restrictions on partitioning:
\begin{enumerate}
\item $\rho$ represents the state of an arbitrary number, $N$, of ensemble
members (see Section~\ref{sec:rule}).
\item Subsets are therefore also of arbitrary size.
\item One cannot explicitly address the number of ensemble members
simultaneously affected by a single, uncontrolled environment.
\item One cannot give individual ensemble members a physically meaningful label.
\end{enumerate}
These restrictions seem to prevent one from avoiding partitioning by
defining within a standard reduced dynamics
any single, ``effective'' environment or any interaction operator that 
\textit{dynamically} mimics the \textit{kinematic}
effects of indistinguishability for the average ensemble member.
In the next two sections, we will use the framework to treat
different physical systems.
To isolate the effects of kinematics, in Section~\ref{sec:bs}
we create a model with trivial dynamics, and in Section~\ref{sec:ap}
we create a model with simple dynamics.

\section{Application:  Beam Splitter}
\label{sec:bs}
\begin{figure}[t]
\includegraphics[trim = 5.1cm 20.1cm 6.4cm 4.4cm, clip, width=.45\textwidth]{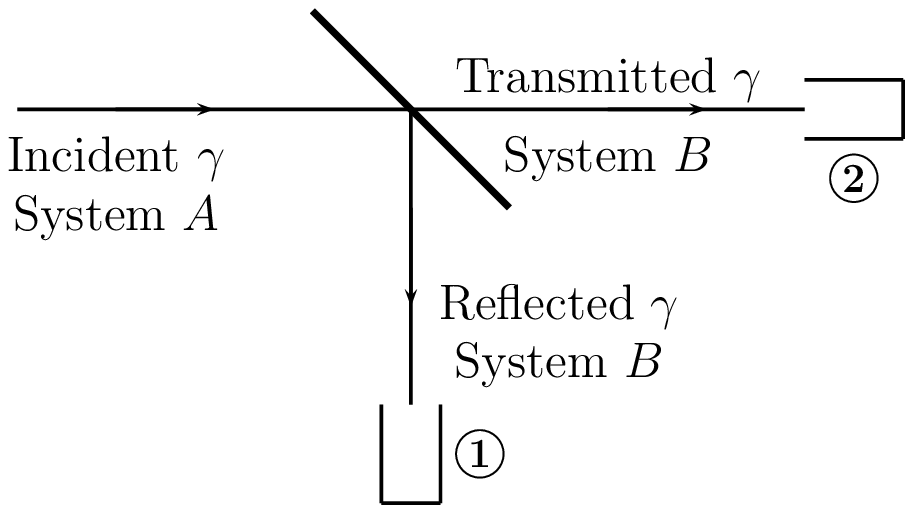}
\caption{Schematic of the photon scattering experiment.
It is divided into System $A$, the incident photons,
and System $B$, the reflected and transmitted photons.
Reflected photons are registered by phototube 1, and
transmitted photons are registered by phototube 2.
Not shown is an imagined source incident from the top.}
\label{fig:schematic}
\end{figure}

Consider the simple system of a photon, $\gamma$,
incident on a beam splitter, being either transmitted into one phototube or else
reflected into a different phototube.
Figure~\ref{fig:schematic} is a schematic of the experiment.
The photon scattering system is interesting as the basis of a photon current.
The probabilistic nature of the scattering at the beam splitter is for the
photon current a source of noise~\cite{buttiker_92} characterized by
correlations between photons in the transmitted and reflected branches.

\subsection{Second quantization calculation}
\label{sec:secquant}
Second quantization provides a simple and effective framework for calculating
correlations, in terms of occupation numbers in the incident,
transmitted, and reflected currents.
In the next section we will use our framework and compare to the 
results of this section.
We are not interested in the microscopic scattering dynamics
for a photon incident on a beam splitter,
so following~\cite{loudon_91} for this simple case
we will deal only with asymptotic in and out states at fixed energy,
and ignore other quantum numbers.

One comment is required, however.
Because the scattering operator, $\hat{S}$, does not change the space of
asymptotic states, one requires another input channel in
Figure~\ref{fig:schematic}, corresponding to photons incident from the top.
Such photons, were they present, would transmit into phototube 1 and reflect
into phototube 2.
This imagined photon source is not present in the actual experiment, however,
so one fixes its input as the vacuum state.
Though it emits no photons,
only when the imagined photon source is present in the calculation
does one calculate the correct values for the correlation
functions~\cite{loudon_91}.

To preserve the symmetry implied by the theoretical approach, 
let us temporarily change the input system to include this imagined source.
The creation and annihilation operators for incoming photons will be
$\hat{a}^\dagger_i$ and $\hat{a}_i$, where $i=1,2$, and
\begin{equation} \label{eq:acom}
[\hat{a}_i,\hat{a}_j^\dagger]=\delta_{ij}.
\end{equation}
The actual source, seen in Figure~\ref{fig:schematic}, will correspond
to $i=1$, and the imagined source will correspond to $i=2$.

For photons in the reflected and transmitted currents, one defines the
creation and annihilation operators, $\hat{b}^\dagger_i$ and
$\hat{b}_i$, with $i=1,2$ corresponding to the
label on the phototubes in Figure~\ref{fig:schematic}, and
also satisfying~\eqref{eq:acom}.

The reflection or transmission at the beam splitter
is then treated as a scattering event, and
one can relate the incoming photons to the transmitted and reflected photons
by~\cite{buttiker_92},
\begin{equation}\label{eq:imscat}
\left(\begin{array}{c}
\hat{b}_1 \\
\hat{b}_2 \\
\end{array}\right) = \left(\begin{array}{cc}
r_{1\,1} & t_{1\,2} \\
t_{2\,1} & r_{2\,2} \\
\end{array}\right) 
\left(\begin{array}{c}
\hat{a}_1 \\
\hat{a}_2 \\
\end{array}\right).
\end{equation}
Here the unitary scattering matrix is written in terms of the
complex reflection coefficient, $r$,
and the complex transmission coefficient, $t$,
which satisfy $\vert r\vert ^2 + \vert t\vert^2 = 1$.
The vacuum state for the input photons is $\ket{0}_1\ket{0}_2\equiv\ket{0}$. 
We will consider an input state, $\ket{n}$,
containing $n$ photons emitted by the actual source, $i=1$, so that
\begin{equation} \label{eq:vacuum}
\ket{n}=\frac{(\hat{a}_1^\dagger)^n}{\sqrt{n!}} \ket{0}.
\end{equation}
Following~\cite{buttiker_92}, we let $\vert r_{11}\vert^2\equiv R$
and $\vert t_{21}\vert^2\equiv T$, and we assume
unit efficiencies for measurement.
The operator representing the occupation number in
the reflected channel of Figure~\ref{fig:schematic}
is $\hat{n}_R=\hat{b}^\dagger_1\hat{b}_1$, and
the the operator representing the occupation number in
the transmitted channel of Figure~\ref{fig:schematic}
is $\hat{n}_T=\hat{b}^\dagger_2\hat{b}_2$.
Using~\eqref{eq:imscat} and~\eqref{eq:vacuum}, one calculates
for the correlations between occupation numbers:
\begin{eqnarray} \label{eq:2quantcor}
\langle \hat{n}_R \hat{n}_T \rangle &=&
\bra{n}\hat{b}^\dagger_1\hat{b}_1\hat{b}^\dagger_2\hat{b}_2\ket{n} = 
R T n(n-1) \nonumber \\
\langle \hat{n}_R \rangle &=& n R \nonumber \\
\langle \hat{n}_T \rangle &=& n T \\
\langle (\Delta \hat{n}_R)^2 \rangle &=& \langle (\Delta \hat{n}_T)^2 \rangle
= n R T \nonumber \\
\langle \Delta \hat{n}_R \Delta \hat{n}_T\rangle &=& - n R T \nonumber
\end{eqnarray}

\subsection{Scattering as an interference event}
\label{sec:opsys}
In this section we will begin to work within the open systems framework
of Section~\ref{sec:mqe}, and in the next section we will introduce losses.
The scattering system of Figure~\ref{fig:schematic} is naturally divided 
at the beam splitter, into subsystems:
\begin{enumerate}
\item \textbf{System $A$, which includes the incident photons:}\\
The density operator $\rho_A(t)$ represents the state of the incident photons.
It is defined in the linear scalar-product space $\Phi_A$, which is
spanned by the ket $\ket{i}$.
In this basis, $\rho_A=\ketbra{i}{i}$.
The Hamiltonian operator is $H_A$, and 
System $B$ is external, so we assume trivial dynamics with
$[\rho_A,H_A]=0$.
\item \textbf{System $B$, which includes
the reflected and transmitted photons:}\\
The density operator for a single photon,
$\rho_B(t)$, is defined in the space $\Phi_B$, which is spanned by the
orthogonal kets $\ket{r}$, representing the state of reflected photons, and 
$\ket{t}$, representing the state of transmitted photons.
The Hamiltonian operator is $H_B$, and the dynamics is again trivial, with
$[\rho_B,H_B]=0$.
\end{enumerate}

To measure the expectation values in~\eqref{eq:2quantcor},
one actively prepares $n$ photons simultaneously in the incident channel.
The $n$ photons then scatter from the beam splitter,
and one measures the state of the $n$ photons
by counting the numbers registered in phototubes 1 and 2.
If one performs $N$ measurements, then one has identically prepared the
state of $n$ photons a total of $N$ times.

Every incident photon is actively 
prepared in a state represented by $\rho_A=\ketbra{i}{i}$.
Eventually every photon reaches the beam splitter and is scattered, which is
the interference event during which the uncontrolled environment passively
prepares every photon such that its state must be described by $\rho_B$.
According to the framework proposed in Section~\ref{sec:mqe},
in addition to the typical reduced
dynamics, density operators undergo the branching
\begin{equation} \label{eq:branching}
\rho_A(t)\longrightarrow w_A\,\rho_A(t+\Delta t) + w_B\,\rho_B(t+\Delta t)
\end{equation}
whenever subsets of ensemble members are passively prepared by an
uncontrolled environment.\footnote{
Though $\rho_A$ and $\rho_B$ are not defined in the same space,
for this scattering system with unit
efficiencies, $w_A=0$, and we can ignore the complication for now.
}
All dynamics for in and out states is trivial,
so we will drop the time dependence from equations in this section.

The scattering matrix in~\eqref{eq:imscat} preserves photon
number~\cite{loudon_91}.
From this we deduce for~\eqref{eq:branching} that $w_A+w_B=1$.
For the simple scattering experiment in which one assumes unit efficiencies,
all photons are eventually scattered into states in System $B$.
We can therefore deduce further that $w_A=0$, fixing $w_B=1$.

If one knows the scattering theory~\eqref{eq:imscat} for the
passive preparation of photons by the beam splitter, then one can write
for $n=1$
\begin{eqnarray} \label{eq:fullent}
\rho_A\longrightarrow \rho_B &=&
            R\, \ketbra{r}{r} + T\, \ketbra{t}{t} \nonumber \\
 & &  + r_{1\,1}t^*_{2\,1}\,\ketbra{r}{t} + 
        t_{2\,1}r^*_{1\,1}\,\ketbra{t}{r}.
\end{eqnarray}
By assuming the most general form for $\rho_B$,
one would also arrive at~\eqref{eq:fullent} up to the coefficients.
For this example with trivial dynamics, the framework simply
reproduces an expression~\eqref{eq:fullent}
commonly used in the study of quantum optics~\cite{gerry_optics_book}.
The only difference for our framework is that~\eqref{eq:fullent}
represents the state of only the $n=1$ subset of the experimental ensemble.

Of course, one often will not know the complete theory for passive preparations
by an unknown environment.
The measurements that lead to~\eqref{eq:2quantcor} are not sensitive to
entanglement, and
the dependence of the correlation functions on $n$ is entirely kinematic.

Without more sophisticated measurements or a complete scattering theory,
by imagining that a physicist actively
prepares any number of photons in a sequence $n_j$, $j=1,2,3\ldots$,
one can deduce a form for $\rho_B$ that, though incomplete, is consistent
with~\eqref{eq:2quantcor}.
For any sequence $n_j$,
the probability that any given photon is reflected is
\begin{equation*}
\sum_j\left(\frac{\langle \hat{n_j}_R\rangle}{n_j}\right) =R
= \mathcal{P}_R(\rho_B).
\end{equation*}
After an incident photon is scattered, then,
the probability to find it in the reflected channel is
\begin{equation} \label{eq:reflprob}
\mathcal{P}_R(\rho_B)=\mathrm{Tr}\left(\ketbra{r}{r}\,\rho_B\right)=R.
\end{equation}

Similarly, the probability to find the photon in the transmitted channel is 
\begin{equation} \label{eq:tranprob}
\mathcal{P}_T(\rho_B)=\mathrm{Tr}\left(\ketbra{t}{t}\,\rho_B\right)=T.
\end{equation}
Knowing~\eqref{eq:reflprob} and~\eqref{eq:tranprob} from measurements,
and without knowing about the entanglement in~\eqref{eq:fullent},
one deduces for~\eqref{eq:branching}
\begin{equation} \label{eq:specbr}
\rho_A\longrightarrow \rho_B=R\, \ketbra{r}{r} + T\, \ketbra{t}{t}.
\end{equation}

We are interested in exploring the kinematics of indistinguishability
without knowing the number, $n$, of photons present, and
therefore not concerned with entanglement.
If the beam splitter is to represent an unknown environment, 
when developing models one presumably would not even know about entanglement
or the imagined photon source.
For the remainder of this section we will ignore off-diagonal
elements of the density operator~\eqref{eq:fullent}
and restrict ourselves to operators of the form~\eqref{eq:specbr}.

In~\eqref{eq:specbr} we have also ignored the kinematics associated with the
simultaneous presence of $n$ incident photons.
One cannot start with~\eqref{eq:specbr} and
directly calculate the correlation functions in~\eqref{eq:2quantcor}.
One way to include the kinematic information is to write the analogue
of~\eqref{eq:specbr} for $n$ incident photons.
This is demonstrated in Appendix~\ref{app:pho} for $n=3$.
Here we will show for general $n$ that by counting indistinguishable
reflection events and indistinguishable transmission events,
one can instead partition the 
experimental ensemble represented by $\rho_B$ in~\eqref{eq:specbr}
into an ensemble of $n$-photon objects while respecting permutation symmetry
and the requirements of Section~\ref{sec:partind}.

Each of the single photons is passively prepared
with probability $R$ to be in the state $\ket{r}$ and with probability
$T$ to be in the state $\ket{t}$.
The ensemble members are otherwise indistinguishable, so to partition the
ensemble we will use the binomial distribution,
\begin{equation*}
\binom{n}{k} R^k T^{n-k},
\end{equation*}
to average over the partitions containing combinations of $k$ 
indistinguishable photons reflected with probability $R$ and
$(n-k)$ indistinguishable photons transmitted with probability $T$.

The expectation values $\langle \hat{n}_R\rangle$ and
$\langle \hat{n}_T\rangle$
are then related to the first moment of the binomial distribution.
We calculate
\begin{equation} \label{eq:first}
\langle \hat{n}_R\rangle = \sum_{k=0}^n \binom{n}{k} R^k T^{n-k} k = n R,
\end{equation}
and
\begin{equation}
\langle \hat{n}_T\rangle = \sum_{k=0}^n \binom{n}{k} R^k T^{n-k} (n-k) = n T.
\end{equation}

In this method, we put $k$ inside the summation to represent the number
of photons reflected.
We put $(n-k)$ inside the summation to represent the number of photons
transmitted.
Therefore $\langle \hat{n}_R \hat{n}_T\rangle$ is
\begin{widetext}
\begin{eqnarray}
\langle \hat{n}_R \hat{n}_T\rangle&=&\sum_{k=0}^n \binom{n}{k} R^k T^{n-k}
k (n-k) = R T n(n-1).
\end{eqnarray}
For the remaining relations, we find
\begin{eqnarray}
\langle (\Delta \hat{n}_R)^2 \rangle &=& \sum_{k=0}^n \binom{n}{k} R^k T^{n-k}
(k-nR)^2 = n R T \\
\langle (\Delta \hat{n}_T)^2 \rangle &=& \sum_{k=0}^n \binom{n}{k} R^k T^{n-k}
\big((n-k)-nT\big)^2 =  n R T \\
\label{eq:last}
\langle \Delta \hat{n}_R \Delta \hat{n}_T\rangle &=& 
\sum_{k=0}^n \binom{n}{k} R^k T^{n-k} 
(k-nR)\big((n-k)-nT\big) = -n R T,
\end{eqnarray}
\end{widetext}
as expected from~\eqref{eq:2quantcor}.
Though we have not yet introduced losses or dynamics, we have treated
the scattering event as an interference event, and we have partitioned
an experimental ensemble.

\subsection{Open systems and inefficiencies} \label{sec:bsineff}
With the generalization of the theory of open systems and
the partitioning method of Section~\ref{sec:opsys}, it is straightforward
to introduce losses into the photon scattering system.
We will consider two sources of inefficiency.
The first occurs if photons are scattered toward the phototubes
of Figure~\ref{fig:schematic} but the phototubes fail to register the photons.
Unless one models the detector quantum mechanically, detector
inefficiency does not require an open system treatment.
Let $\Lambda_R^\prime$ be the operator representing observation of
reflected photons,
and let $\Lambda_T^\prime$ represent observation of transmitted photons.
If there is some probability of a failed registration, then one has
\begin{equation*}
\Lambda_R^\prime\Rightarrow(1-\epsilon_R)\ketbra{r}{r}\quad
\textrm{and} \quad
\Lambda_T^\prime\Rightarrow(1-\epsilon_T)\ketbra{t}{t},
\end{equation*}
where $\epsilon_R$ is the probability that a photon fails to register
in phototube 1, and
$\epsilon_T$ is the probability that a photon fails to register
in phototube 2.
The probability to measure a reflected photon is 
\begin{equation*}
\mathcal{P}_R\big(\rho_B\big) = (1-\epsilon_R)R,
\end{equation*}
and the probability to measure a transmitted photon is
\begin{equation*}
\mathcal{P}_T\big(\rho_B\big) = (1-\epsilon_T)T.
\end{equation*}

The second type of inefficiency occurs when the beam splitter scatters photons
into such directions that they will not hit either phototube.
Then one introduces another system, System $C$,
that includes these photons.
System $B$ is the system of interest, and
System $C$ corresponds to the environmental system of the typical
framework~\cite{petruccione_open_quantum_systems}.
Unmeasurable photons are in a state represented by a density operator,
$\rho_C$, constructed from vectors in the space $\Phi_C$,
which does not contain $\ket{r}$ or $\ket{t}$.
Instead of~\eqref{eq:specbr}, one uses
\begin{eqnarray} \label{eq:paspc}
\rho_A&\longrightarrow& w_B\,\rho_B + w_C\,\rho_C \nonumber \\
 & \, &= w_B\big(R\ketbra{r}{r}+T\ketbra{t}{t}\big) + w_C\,\rho_C
\end{eqnarray}
to represent passive preparation at the beam splitter
(again ignoring entanglement).

It is important to note again that~\eqref{eq:paspc}
is not mathematically correct
because $\rho_B$ and $\rho_C$ are not defined in the same space.
The probability after scattering to measure
the observable represented by $\Lambda$ is
\begin{eqnarray}
\mathcal{P}_{\Lambda_B}\big(w_B\rho_B\big)+
\mathcal{P}_{\Lambda_C}\big(w_C\rho_C\big)\qquad  \nonumber \\
= w_B\mathcal{P}_{\Lambda_B}\big(\rho_B\big)+
w_C\mathcal{P}_{\Lambda_C}\big(\rho_C\big),
\end{eqnarray}
where $w_B+w_C=1$, and
$\Lambda_C$ is the appropriate operator in the space $\Phi_C$.
Because System $C$ contains photons that are not measured, $\Lambda_C$ is
the zero operator in $\Phi_C$.
The probability to measure a reflected photon is 
\begin{equation*}
\mathcal{P}_R\big(w_B \rho_B\big) = w_B R,
\end{equation*}
and the probability to measure a transmitted photon is
\begin{equation*}
\mathcal{P}_T\big(w_B \rho_B\big) = w_B T.
\end{equation*}

In real experiments, both sources of measurement inefficiency will appear,
and one will have for the probabilities
\begin{equation*}
\mathcal{P}_R\big(\rho_B\big) = w_B(1-\epsilon_R) R
\end{equation*}
and
\begin{equation*}
\mathcal{P}_T\big(\rho_B\big) = w_B(1-\epsilon_T) T.
\end{equation*}

Now we can check the results in~\eqref{eq:2quantcor}, for the case
that one does not have unit efficiencies.
Let $R^\prime$ be the probability that a reflected photon is registered
and $T^\prime$ be the probability that a transmitted photon is registered.
Let $\eta$ be the probability that an ensemble member is not measured,
such that 
\begin{equation} \label{eq:nrm}
R^\prime + T^\prime + \eta = 1.
\end{equation}
If the experimental ensemble contains \emph{all} photons actively prepared
in the incident channel, then
\begin{eqnarray} \label{eq:param1}
R^\prime&=&w_B(1-\epsilon_R)R \nonumber \\
T^\prime&=&w_B(1-\epsilon_T)T \\
\eta&=&1-w_B(1-\epsilon_R R-\epsilon_T T). \nonumber
\end{eqnarray}

If, instead, the experimental ensemble contains only those photons scattered
into System~$B$, then $w_B = 1$, and
\begin{eqnarray} \label{eq:param2}
R^\prime&=&(1-\epsilon_R)R \nonumber \\
T^\prime&=&(1-\epsilon_T)T \\
\eta&=&\epsilon_R R+\epsilon_T T. \nonumber
\end{eqnarray}

In general, one will not know if a photon remains unmeasured because of
inefficiencies in scattering or because of inefficiencies in the phototubes.
In other words, one can measure $R^\prime$ and $T^\prime$, but one will 
not know which of~\eqref{eq:param1} or~\eqref{eq:param2} is appropriate.

In~\eqref{eq:paspc} we have the theory for the state of any single photon
before and after an interference event.
To calculate correlation functions for different $n$, 
again we partition the experimental ensemble.
Now we must use the multinomial distribution:
\begin{widetext}
\begin{eqnarray}
\label{eq:meqfirst} \langle \hat{n}_R \rangle&=&
\sum_{j=0}^n \sum_{k=0}^n \sum_{\ell=0}^n
\frac{n!}{j!k!\ell !} R^{\prime j} T^{\prime k} \eta^{\prime \ell}
\,\delta_{n,j+k+\ell}\, j
= n R^{\prime} (R^\prime + T^\prime + \eta)^{n-1} \\
\langle \hat{n}_T \rangle&=&
\sum_{j=0}^n \sum_{k=0}^n \sum_{\ell=0}^n
\frac{n!}{j!k!\ell !} R^{\prime j} T^{\prime k} \eta^{\prime \ell}
\,\delta_{n,j+k+\ell}\, k
= n T^{\prime} (R^\prime + T^\prime + \eta)^{n-1} \\
\langle \hat{n}_R\hat{n}_T \rangle&=&
\sum_{j=0}^n \sum_{k=0}^n \sum_{\ell=0}^n
\frac{n!}{j!k!\ell !} R^{\prime j} T^{\prime k} \eta^{\prime \ell}
\,\delta_{n,j+k+\ell}\, j k 
=   R^\prime T^{\prime}n(n-1) (R^\prime + T^\prime + \eta)^{n-2}
\end{eqnarray}
where $\delta_{n,j+k+\ell}$ is the Kronecker delta.
Because of~\eqref{eq:nrm}, we have
$\langle \hat{n}_R \rangle=n R^\prime$,
$\langle \hat{n}_T \rangle=n T^\prime$, and
$\langle \hat{n}_R\hat{n}_T \rangle= R^\prime T^\prime n (n-1)$.

Using~\eqref{eq:nrm}, the other expectation values are
\begin{eqnarray}
\langle(\Delta \hat{n}_R)^2 \rangle&=&
\sum_{j=0}^n \sum_{k=0}^n \sum_{\ell=0}^n
\frac{n!}{j!k!\ell !} R^{\prime j} T^{\prime k} \eta^{\prime \ell}
\,\delta_{n,j+k+\ell}\, (j-n R^\prime)^2 
= n R^\prime (T^\prime+\eta) \\
\langle(\Delta \hat{n}_T)^2 \rangle&=&
\sum_{j=0}^n \sum_{k=0}^n \sum_{\ell=0}^n
\frac{n!}{j!k!\ell !} R^{\prime j} T^{\prime k} \eta^{\prime \ell}
\,\delta_{n,j+k+\ell}\, (k-n T^\prime)^2 
= n T^\prime (R^\prime+\eta) \\
\langle\Delta \hat{n}_R\Delta \hat{n}_T \rangle&=&
\sum_{j=0}^n \sum_{k=0}^n \sum_{\ell=0}^n
\frac{n!}{j!k!\ell !} R^{\prime j} T^{\prime k} \eta^{\prime \ell}
\,\delta_{n,j+k+\ell}\, (j-n R^\prime)(k-n T^\prime) 
= -n R^\prime T^\prime
\label{eq:meqlast}
\end{eqnarray}
\end{widetext}

Comparing~\eqref{eq:meqfirst}--\eqref{eq:meqlast}, where
$R^\prime + T^\prime + \eta = 1$, one can see how to generalize 
to imperfect measurements the
results~\eqref{eq:2quantcor} of Section~\ref{sec:secquant}, 
where $R+T=1$ and inefficiencies were ignored.
The main difference is that
$\langle(\Delta \hat{n}_R)^2 \rangle \neq \langle(\Delta \hat{n}_T)^2 \rangle$
when $\eta\neq 0$.
If one writes from~\eqref{eq:2quantcor} that
$\langle(\Delta \hat{n}_R)^2 \rangle = n R (1-R)$ and
$\langle(\Delta \hat{n}_T)^2 \rangle = n T (1-T)$, however,
then one can correctly take $R\rightarrow R^\prime$ and
$T\rightarrow T^\prime$.

In the language of the theory of open
systems~\cite{petruccione_open_quantum_systems},
the system of interest includes measurable photons in states described by
$\rho_B$, which is constructed from vectors in $\Phi_B$.
The environment contains unmeasurable photons in states described by
$\rho_C$, which is constructed from vectors in $\Phi_C$.
All scattered photons are
represented by vectors in $\Phi=\Phi_B\oplus\Phi_C$,
which is the space of states of the combined ``system plus environment,''
when entanglement with the environment is ignored.

In scattering experiments, one does not measure a dynamical dissipation in
currents.
In~\eqref{eq:specbr} and~\eqref{eq:paspc} we therefore
have models without dynamics.
Inefficiencies appear as a measurement event rate smaller than
the incidence rate per area of incoming objects, for a given cross section.
While an experimenter continues to supply photons, the losses appear only as
measurement inefficiencies, which are treated by normalizing measured data
to the total number of photons registered in the phototubes over the course
of the experiment.

\section{Application:  Rabi Oscillations Experiments}
\label{sec:ap}
As an application with simple dynamics,
we will create a toy model for systems undergoing Rabi oscillations.
Here time cannot be ignored, and we must deduce a form for
environmental interactions.
We will also implement the technique used in the previous
section to partition the ensemble of single photons
represented by~\eqref{eq:specbr} into an ensemble of $n$-photon objects.

\subsection{Rabi Oscillations}
The theory of Rabi oscillations is well established~\cite{dodd_atoms_1991}.
It describes a two level system, with levels described by
the projection operators
$\ket{g}\bra{g}$ representing the ground state and
$\ket{e}\bra{e}$ representing the excited state.
When the system is prepared at $t=0$ to be in the excited state,
$\ket{e}$, and it is coupled to a correctly tuned
radiation field,
the probability at $t$ to find the system in the ground state,
$\ket{g}$, is
\begin{equation} \label{born_rabi}
\mathcal{P}_{\ket{g}\bra{g}}\big(\rho(t)\big)= \textrm{sin}^2(\Omega t)=
\frac{1}{2}\big(1-\textrm{cos}(2 \Omega t)\big),
\end{equation}
where $\Omega$ is called the Rabi frequency.
(The probability to find the system in the same state in which it was prepared
is $\textrm{cos}^2(\Omega t)$.)
These expressions are calculated using the dynamics for closed systems 
undergoing Rabi oscillations.
The probability in~\eqref{born_rabi} is to be compared with experimental
results~\eqref{eq:comparison}.

In actual experiments, because of environmental interference
one never measures the result in~\eqref{born_rabi}.
In this paper we wish to consider several very different
but very clean and well-controlled
experiments~\cite{meekhof_rabi_1996,brune_rabi_1996,petta_coherent_2005,
cole_nature_2001,zrenner_nature_2002,wang_macdonald_prb_2005,ramsay_prl_2010}.

In one experiment~\cite{meekhof_rabi_1996},
also discussed in Section~\ref{sec:meas},
internal levels of a $^9$Be$^+$ ion couple to the harmonic binding potential.
Rabi oscillations occur between two of the coupled internal and vibrational
levels, and the oscillations between different sets of levels are measured.
In another experiment~\cite{brune_rabi_1996},
Rabi oscillations are observed between the
circular states of a Rydberg atom coupled to a field stored in
a high $Q$ cavity.
In~\cite{petta_coherent_2005},
the Rabi oscillations are between the spin states of
two electrons in a double quantum dot.
In~\cite{cole_nature_2001}, Rabi oscillations occur between motional states
of electrons bound to shallow donors in semiconductors.
These motional states mimic the levels of a hydrogen atom.
In three experiments~\cite{zrenner_nature_2002,wang_macdonald_prb_2005,
ramsay_prl_2010}, oscillations occur between excitation levels 
of electrons confined in quantum dots.

The measured results
in~\cite{meekhof_rabi_1996,brune_rabi_1996,petta_coherent_2005},
do not match the prediction in~\eqref{born_rabi}.
Instead, the measured probability is fit by an appropriately damped sinusoid
that is a function of the time duration, $\tilde{t}$, from
active preparation,
\begin{equation}
\label{damped_rabi}
\mathcal{P}_{\ket{g}\bra{g}}\big(\rho(\tilde{t})\big)=
\frac{1}{2}\big(1-e^{-\gamma \tilde{t}}\,\textrm{cos}(2 \Omega \tilde{t})\big),
\end{equation}
where $\gamma$ is an experimentally determined damping factor.

In~\cite{petta_coherent_2005}, amplitude, offset, and phase are fit as well,
because the oscillations are measured also as a function of
a swept detuning voltage.
In~\cite{wang_macdonald_prb_2005,ramsay_prl_2010}, photocurrent is measured, and
from it the dephasing of the Rabi oscillations is inferred.
Numerically solving the models they fit to their data,
the subsystems undergoing Rabi oscillations fit~\eqref{damped_rabi}.

\subsection{Dynamics for the toy model}
\label{sec:model}
Already having the kinematic requirements for open systems and indistinguishable
ensemble members, our next task is to deduce the proper dynamics
for the experimental system in question.
This means we must identify the spaces $\Phi_i$ and the
corresponding algebras of observables.
Starting with the branched density operator from~\eqref{eq:cases},
we calculate the probability at a duration $\tilde{t}$ from
active preparation:
\begin{eqnarray}
\label{eq:long_prob}
\mathcal{P}_{\Lambda}\big(\rho(\tilde{t})\big) & = &
\Tr \big( \Lambda_0 \,a_0\, \rho_0(t_0)\big) +
\Tr \big( \Lambda_1 \,a_1\, \rho_1(t_1)\big) + \ldots  \nonumber \\
& = & a_0\,\mathcal{P}_{\Lambda_0}\big(\rho_0(t_0)\big) +
      a_1\,\mathcal{P}_{\Lambda_1}\big(\rho_1(t_1)\big) + \ldots \,\,\,\,\,\,
\end{eqnarray}
where, as before, $\rho_i$ is constructed from vectors in $\Phi_i$.

For
the long-time behavior of the measured result in~\eqref{damped_rabi}, one has
\begin{equation*}
\lim_{\tilde{t}\rightarrow\infty}
\mathcal{P}_{\ket{g}\bra{g}}\big(\rho(\tilde{t})\big)=
\frac{1}{2}.
\end{equation*}
From this we deduce that there is no dissipation for the
experimental system,
and that the terms $\mathcal{P}_{\Lambda_i}\big(\rho_i(t_i)\big)$
in~\eqref{eq:long_prob} ought not to introduce the type of losses
considered for the beam splitter in Section~\ref{sec:bsineff}.
For these experiments, we then hypothesize that 
immediately after the interference event, $\Phi_0=\Phi_1=\Phi_i$
for all $i$, so we can take $\Phi_i\rightarrow\Phi$.
The algebra of observables will therefore be the typical algebra for the closed
system, so we will take $\Lambda_i\rightarrow\Lambda=\ket{g}\bra{g}$.
We want a probability of the form
\begin{equation} \label{eq:sum_branch}
\mathcal{P}_{\Lambda}\big(\rho(\tilde{t})\big) =
\sum_{m=0}^{M_{branch}}\,
      a_m\,\mathcal{P}_\Lambda\big(\rho_m(t_m)\big)
\end{equation}
where $M_{branch}$ is the number of branches.
To keep track of the branches, we will find it convenient to retain the
label $i$ for the $\rho_i$, even though all $\rho_i$ are defined in the
same space, $\Phi$.
This formula is deceptively simple in appearance, because, in general,
the number of branches, $M_{branch}$,
will depend on $\tilde{t}$, and the value of $a_{m^\prime}$ 
for a fixed $m^\prime$ will depend on $M_{branch}$.

The $\rho_i$ are all defined in $\Phi$, and to keep our model simple
we will assume that during interference events
the environment prepares the system to be in energy eigenstates, 
$\ket{g}$ and $\ket{e}$, ignoring entanglement.
When a passive preparation occurs
at $t_i=0$, the most general form for $\rho_i$ in~\eqref{eq:cases} is then
\begin{equation}
\label{eq:garb1}
\rho_i(t_i=0)=w_i\,\,\ket{g}\bra{g} + v_i\,\, \ket{e}\bra{e},
\end{equation}
with $w_i+v_i=1$.
In~\eqref{eq:garb1}, $w$ is a weight appropriate for ensemble members 
prepared passively to be in the ground state, and $v$ is a weight for 
ensemble members prepared in the excited state.

Note that this is not a new idea.
Leggett has suggested~\cite{leggett_arrow_qmeasurement} that decoherence
``is exactly the result of a `measurement' whose result is uninspected,''
and he has called this process ``garbling.''
With this toy model, developed within our framework, we will
treat the effects of garbling.
We ignore the dynamics of the measurement process itself
because it is simply not understood
and apparently occurs over very small durations in time.

\subsection{Born probability}
\label{sec:prob_indist}
For simplicity, we will assume that the passive measurements for 
our systems are fair measurements, though this would not generally
be required.
To isolate kinematic effects, we will also ignore any reduced dynamics
expressed in~\eqref{eq:redtime} and~\eqref{eq:type1sum}.

When modeling experiments or using more sophisticated models for
environmental interaction, one would probably use a full simulation.
To maintain transparency, however, we will search for an analytical formula
for our toy model.
We will start with a sequence of interference events described in
Section~\ref{sec:ievents}.
Because the damping of the oscillations in~\eqref{damped_rabi}
is exponential, the decay rate is constant, and
we can deduce that there should be a single timescale,
$\Delta\tilde{t}$, for regularly spaced interference events.
Then the Born probability when there may have been $n$ interference events is
$\mathcal{P}_{\ket{g}\bra{g}}\big(\rho(n\Delta\tilde{t})\big)$.

As discussed in Section~\ref{sec:rule}, models should not depend on
the ensemble size or on the number of members simultaneously present.
Intuitively, one would like a model that does depend on the number of
interference events that have occurred.
Note from our treatment of photon scattering that counting indistinguishable
ensemble members is equivalent to counting as indistinguishable the events
in which they are prepared.

We are consequently led to a model with two parameters.
$\Delta\tilde{t}$ is the timescale for interference.
$\beta$ represents the probability that a randomly chosen
time interval will have preceded an interference event,
with $0\le \beta\leq 1$ and $\beta=1$ for a perfectly isolated system.
This can be better understood after equation \eqref{eq:mom}.
We will again use the binomial distribution,
\begin{equation*}
b(n,k,\beta) \equiv
{n \choose k} \beta^k (1-\beta)^{n-k},
\end{equation*}
and the normalization
\begin{equation*}
\sum_{k=0}^{n} {n \choose k} \beta^k (1-\beta)^{n-k} = 1.
\end{equation*}

To illustrate the model, let us write~\eqref{eq:long_prob} for
the case that the average ensemble member will have suffered only one
interference event before an active measurement occurs.
This will truncate our formula at a reasonable size, and we will generalize
to multiple events below.
The probability at $4\Delta\tilde{t}$
to find in $\ket{g}$ members that have been initially
prepared at  $\tilde{t}=0$ to be in $\ket{e}$ is
\begin{widetext}
\begin{eqnarray}
\label{eq_simple_case}
\mathcal{P}_{\ket{g}\bra{g}}\big(\rho(4\Delta\tilde{t})\big) & = &
    b(4,4,\beta)\Big(\textrm{sin}^2(\Omega\, 4\Delta\tilde{t})\,\textrm{cos}^2(\Omega\, 0\Delta\tilde{t})
 + \textrm{cos}^2(\Omega\, 4\Delta\tilde{t})\,\textrm{sin}^2(\Omega\, 0\Delta\tilde{t})\Big)+\nonumber \\
& & b(4,3,\beta)\Big(\textrm{sin}^2(\Omega \,3\Delta\tilde{t})\,\textrm{cos}^2(\Omega \,1\Delta\tilde{t})
 + \textrm{cos}^2(\Omega \,3\Delta\tilde{t})\,\textrm{sin}^2(\Omega \,1\Delta\tilde{t})\Big)+\nonumber \\
& & b(4,2,\beta)\Big(\textrm{sin}^2(\Omega \,2\Delta\tilde{t})\,\textrm{cos}^2(\Omega \,2\Delta\tilde{t})
 + \textrm{cos}^2(\Omega \,2\Delta\tilde{t})\,\textrm{sin}^2(\Omega \,2\Delta\tilde{t})\Big)+\nonumber \\
& & b(4,1,\beta)\Big(\textrm{sin}^2(\Omega \,1\Delta\tilde{t})\,\textrm{cos}^2(\Omega \,3\Delta\tilde{t})
 + \textrm{cos}^2(\Omega \,1\Delta\tilde{t})\,\textrm{sin}^2(\Omega \,3\Delta\tilde{t})\Big)+\nonumber \\
& & b(4,0,\beta)\Big(\textrm{sin}^2(\Omega \,0\Delta\tilde{t})\,\textrm{cos}^2(\Omega \,4\Delta\tilde{t})
 + \textrm{cos}^2(\Omega \,0\Delta\tilde{t})\,\textrm{sin}^2(\Omega \,4\Delta\tilde{t})\Big).
\end{eqnarray}
\end{widetext}

The explanation of \eqref{eq_simple_case} is straightforward.
It is in the form of the probability in~\eqref{eq:long_prob}, and each line
corresponds to a branch of the density operator in~\eqref{eq:cases}.
From partitioning the ensemble for the beam splitter (Section~\ref{sec:bs}),
we know that the size of a subset is proportional to the number
of events in which ensemble members are appropriately prepared,
if we count the events themselves as indistinguishable.
We must relate the binomial distribution to the passage of time, so
at every step $k$, with $0\leq k\leq n=4$, we count the 
(normalized) number of combinations for arranging the $n$
time intervals between possible event times,
such that $k$ of them came before the single interference event.
This number is given by $b(n,k,\beta)$, and is called $a_i$
in~\eqref{eq:long_prob} and~\eqref{eq:cases}.

Because possible interference events occur at increments of the time
scale, $\Delta\tilde{t}$, the weight $b(n,k,\beta)$ is attached
to any passive preparation occurring at $k\Delta\tilde{t}$.
Consider the second line, where $k=3$, and compare
to the notation in~\eqref{eq:garb1}.
The weight $w=\textrm{sin}^2(\Omega \,3\Delta\tilde{t})$.
The weight $v=\textrm{cos}^2(\Omega \,3\Delta\tilde{t})$.
The duration from passive preparation to $4\Delta\tilde{t}$ is then
$(4-3)\Delta\tilde{t}=1\Delta\tilde{t}$, so the functions with argument
$1\Delta\tilde{t}$ contain the time dependence of the Born probability
of the mixed state.

Let us introduce the notation
$\mathcal{P}^{(i)}_{\ket{g}\bra{g}}\big(\rho(n\Delta\tilde{t})\big)$
to represent the Born probability under the assumption that members
on average will have suffered $i$ interference events before active measurement.
Then for general $n$,
\begin{widetext}
\begin{equation}
\label{eq:one_event}
\mathcal{P}^{(1)}_{\ket{g}\bra{g}}\big(\rho(n\Delta\tilde{t})\big) =
\sum_{k=0}^n b(n,k,\beta)
\Big(\textrm{sin}^2(\Omega\,k \Delta\tilde{t})\,\textrm{cos}^2(\Omega(n-k)\Delta\tilde{t})
 + \textrm{cos}^2(\Omega\,k\Delta\tilde{t})\,\textrm{sin}^2(\Omega(n-k)\Delta\tilde{t})\Big).
\end{equation}
By simply exchanging
$\textrm{cos}^2(\Omega\, k\Delta\tilde{t})$ and
$\textrm{sin}^2(\Omega\, k\Delta\tilde{t})$,
we calculate
$\mathcal{P}^{(1)}_{\ket{e}\bra{e}}\big(\rho(n\Delta\tilde{t})\big)$,
which is the probability to find the ensemble members
in the excited state.

In \eqref{eq:one_event},
we have assumed that ensemble members will have suffered one
interference event.
To allow for the possibility of multiple events, the terms
$\textrm{sin}^2(\Omega\, k\Delta\tilde{t})$ and
$\textrm{cos}^2(\Omega\, k\Delta\tilde{t})$ 
must be replaced with new functions of $k\Delta\tilde{t}$,
that predict the effects of interference events prior to the
single event assumed in \eqref{eq:one_event}.
For general $i$, we get the recursive expression
\begin{equation}
\label{eq:two_events}
\mathcal{P}^{(i)}_{\ket{g}\bra{g}}\big(\rho(n\Delta\tilde{t})\big) = \sum_{k=0}^n b(n,k,\beta)
\Big(\mathcal{P}^{(i-1)}_{\ket{g}\bra{g}}\big(\rho(k\Delta\tilde{t})\big)\,
\textrm{cos}^2(\Omega(n-k)\Delta\tilde{t})
 + \mathcal{P}^{(i-1)}_{\ket{e}\bra{e}}\big(\rho(k\Delta\tilde{t})\big)\,
\textrm{sin}^2(\Omega(n-k)\Delta\tilde{t})
\Big).
\end{equation}
\end{widetext}

We have written the Born probability as a function of $n\Delta\tilde{t}$.
The final step is to scale our result back to the continuous $\tilde{t}$.
The first moment of the binomial distribution is
\begin{equation}
\label{eq:mom}
\langle k \rangle =
\sum_{k=0}^{n} {n \choose k} \beta^k (1-\beta)^{n-k} \,k =\beta\, n.
\end{equation}
After stepping through time to $n\Delta\tilde{t}$,
on average $\beta\,n$ of the intervals will have preceded
the interference event number $i$.
This provides us a physical interpretation of our two parameters.
To ensure that the time scales with something physical, as indicated
by the exponential nature of the damping,
and scales as well with the recursive index $i$, we will need to use
$\langle k \rangle \Delta\tilde{t}= \beta\, n\Delta\tilde{t}=\tilde{t}$.
After a calculation of the probability as a function of
$n\Delta\tilde{t}$, we make the replacement
\begin{equation} \label{eq:trescale}
n\rightarrow \frac{\tilde{t}}{\beta \,\Delta\tilde{t}}\,.
\end{equation}
This restricts us to non-zero values of $\beta$ and $\Delta\tilde{t}$.
We have also simply interpolated between the discrete
values of $n$ at which \eqref{eq:two_events} is actually defined.
Because our time scale is understood to be an average value,
it would be inappropriate to assume for our model anything more complicated.

\begin{figure}[h]
\includegraphics[width=.5\textwidth]{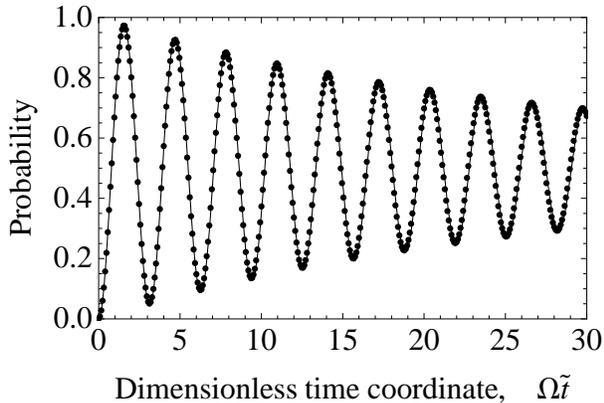}
\caption{Plot of the Born probability,
$\mathcal{P}^{(5)}_{\ket{g}\bra{g}}\big(\rho(\tilde{t})\big)$,
for indistinguishable ensemble members.
For the dots we have used equations \eqref{eq:two_events}
and \eqref{eq:subst}.
The solid line is a plot of the damped sinusoid \eqref{damped_rabi}
that fits the experimental data.
We have used $\Omega\Delta\tilde{t}\approx 0.7$ and $\beta=0.995$,
and we have fit $\gamma / \Omega = 0.039$.}
\label{fig:indis}
\end{figure}

The probability at the duration, $\tilde{t}$, from active preparation
and assuming on average $i$ interference events, is therefore
\begin{equation}
\label{eq:subst}
\mathcal{P}^{(i)}_{\ket{g}\bra{g}}\big(\rho(\tilde{t})\big)=
\mathcal{P}^{(i)}_{\ket{g}\bra{g}}\big(\rho(n\Delta\tilde{t})\big),\quad
n\rightarrow\frac{\tilde{t}}{\beta\,\Delta\tilde{t}}\,.
\end{equation}
In Figure \ref{fig:indis} we have plotted 
$\mathcal{P}^{(5)}_{\ket{g}\bra{g}}\big(\rho(\tilde{t})\big)$.
For clean experiments we expect a good approximation for $i=5$.
The agreement with the experimentally measured damped sinusoid is very good.

\subsection{Measurable consequence of indistinguishability}
\label{sec:meascons}
Experiments reveal that the damping factor, $\gamma$, seems to depend
generally on the Rabi frequency, $\Omega$, regardless of the nature of the
experiment.
In the experiment with $^9\textrm{Be}^+$ ions~\cite{meekhof_rabi_1996},
the different levels are described by the kets
$\ket{\downarrow, n}$ and $\ket{\uparrow, n+1}$, where
$\ket{\downarrow}$ and $\ket{\uparrow}$ are internal states of the Be ion, and
$\ket{n}$ represents vibrational Fock states.
Rabi oscillations are measured for the
frequencies~\cite{wineland_experimental_1998,meekhof_rabi_1996}
\begin{equation} \label{eq:freq_seq}
\Omega_{n,n+1} =
\Omega \frac{0.202 \,e^{-0.202^2/2}}{\sqrt{n+1}} L^1_n(0.202^2),
\end{equation}
where $L^1_n$ is the generalized Laguerre polynomial.
The corresponding damping factor, $\gamma_n$, is measured
to increase with $n$ according to 
\begin{equation}
\label{eqn:gamma_meas}
\frac{\gamma_n}{\gamma_0}\approx (1+n)^{0.7}. \qquad \textrm{(Measured)}
\end{equation}

In the experiment~\cite{petta_coherent_2005}
with spin states of two electrons in a double quantum dot,
the damping factor, $\gamma$, is stated to be proportional to the
Rabi frequency, $\Omega$, though no mathematical relation is given.

For~\cite{cole_nature_2001,zrenner_nature_2002,
wang_macdonald_prb_2005,ramsay_prl_2010},
this dependence has been called Excitation Induced Dephasing (EID),
and its cause has remained an open question.
In~\cite{wang_macdonald_prb_2005,ramsay_prl_2010}, the dependence of the
damping factor on the Rabi frequency has been found to be
$\gamma\propto\Omega^2$.

These results for such different experiments
suggest that some dependence of $\gamma$ on $\Omega$ may be due to a
general, kinematic effect rather than any dynamics specific to an experimental
system or its uncontrolled environment.

When we include the kinematics of indistinguishability in our toy model, we
find that the damping factor, $\gamma$, \emph{does in general} depend on the 
Rabi frequency, $\Omega$, in agreement with measurements and the phenomenon
called Excitation Induced Dephasing.
For $\beta$ close to $1$, the probability is dominated by the
terms proportional to $b(n,k=n,\beta)$.
Solving the rather crude, truncated form\footnote{
This would correspond also to
a model for passive preparations always into the ground state.
}
\begin{equation}
\mathcal{P}_{\ket{g}\bra{g}}\big(\rho(n\Delta\tilde{t})\big)
\approx \sum_{k=0}^n
b(n,k,\beta)\textrm{sin}^2(\Omega\,k\Delta\tilde{t})
\end{equation}
and using \eqref{eq:subst}, we get
\begin{eqnarray}
\mathcal{P}_{\ket{g}\bra{g}}\big(\rho(\tilde{t})\big)& \approx & 
\frac{1}{4} \Big(2-
\big( 1-\beta (1-e^{-2 i \Delta\tilde{t}\Omega})\big)^\frac{\tilde{t}}{\beta\Delta\tilde{t}}
\nonumber \\
\label{eq:approx}
& &\,\,\quad-\big( 1-\beta
(1-e^{+2 i \Delta\tilde{t}\Omega})\big)^\frac{\tilde{t}}{\beta\Delta\tilde{t}}
\Big).\quad
\end{eqnarray}
For small $\Delta\tilde{t}$, \eqref{eq:approx} is
\begin{equation}
\mathcal{P}_{\ket{g}\bra{g}}\big(\rho(\tilde{t})\big)=
\frac{1}{2}\Big(1-e^{-\gamma \tilde{t}}\,\big(\textrm{cos}(2 \Omega \tilde{t})
+\textrm{O}(\Delta\tilde{t}^{\,2})\big)\Big),
\end{equation}
where
\begin{equation}
\gamma = 2\, (1-\beta)\, \Omega^2\, \Delta\tilde{t} + \textrm{O}(\Delta\tilde{t}^{\,3}).
\end{equation}
For $\beta \approx 1$ and $\Omega\Delta\tilde{t} \ll 1$, 
$\gamma$ is quadratic in $\Omega$.
This matches the results in~\cite{wang_macdonald_prb_2005,ramsay_prl_2010}.

\begin{figure}[h]
\includegraphics[width=.5\textwidth]{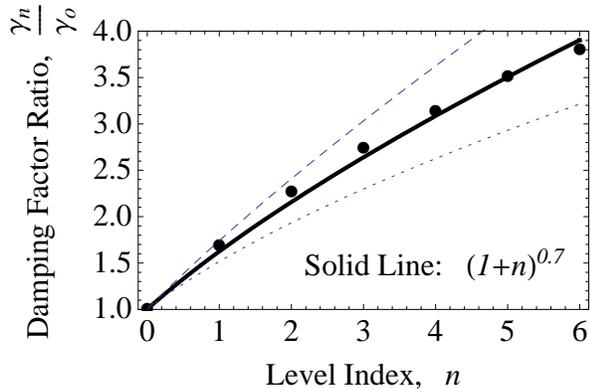}
\caption{Matching the experimental results for the ratio of
damping factors, $\frac{\gamma_n}{\gamma_0}$.
The large dots result from our theoretical calculation of the Born
probability using \eqref{eq:two_events} and \eqref{eq:subst}.
The solid line is the experimentally measured relation, $(1+n)^{0.7}$.
To indicate a scale for the exponent,
the thin dashed line is a plot of $(1+n)^{0.8}$,
and the thin dotted line is a plot of $(1+n)^{0.6}$.}
\label{fig:fit}
\end{figure}

Because we made no assumption regarding the dynamics of environmental
interference,
the same model is general enough to apply also 
to the system in~\cite{meekhof_rabi_1996},
and with it we fit the measured relation \eqref{eqn:gamma_meas}
between $\gamma$ and $\Omega$.
Figure~\ref{fig:fit} is the result of fitting the damping factor,
$\gamma_n$, to 
$\mathcal{P}^{(5)}_{\ket{g}\bra{g}}\big(\rho(\tilde{t})\big)$,
calculated with the sequence of frequencies in \eqref{eq:freq_seq}
and with $\Omega_0\Delta\tilde{t}\approx 0.2$.
(The exponent can be shifted by choosing different time scales.)
Unfortunately, we are limited by computational resources to $n\leq 6$,
and for $n=6$ we were only able to use the first few oscillations when fitting.

Figure~\ref{fig:fit} shows a very good quantitative agreement with
experiment \eqref{eqn:gamma_meas}.
Our study indicates that the decoherence measured in~\cite{meekhof_rabi_1996}
is explained simply by the environmental interference
characterized in Section~\ref{sec:model} and
having a characteristic frequency of $\approx 5\Omega_0$.

In the standard decoherence program, one typically treats these experiments
using the decoherence master equation, and ignores the possibility
that different ensemble members may suffer environmental
interference differently.
The generic solution of the master equation describing a system
undergoing Rabi oscillations~\cite{petruccione_open_quantum_systems} predicts
a constant $\gamma$, with no dependence on $\Omega$.
This is one reason why the measured results have been puzzling.

When assuming that all ensemble members suffer interference identically,
one has not needed to consider the kinematics associated with
indistinguishability.
As a consequence, one has searched for dynamical explanations for the
measured dependence of $\gamma$ on $\Omega$.
For example, for the experiment on Be ions~\cite{meekhof_rabi_1996}, 
one has required models with detailed interaction
terms exclusive to the experiment itself.
Among other
things~\cite{murao_decoherence_1998,bonifacio_pra_2000,difidio_damped_2000},
one has assumed laser amplitude noise~\cite{schneider_decoherence_1998},
decay from intermediate levels and, simultaneously, charge-coupling of the ion
with the trap's
electrodes~\cite{budini_localization_2002,budini_dissipation_2003},
or feedback damping from polarized background gas polarized by the
oscillating ions themselves~\cite{serra_decoherence_2001}.
While plausible, none of these studies has resulted in an agreement as
quantitatively good as that in Figure~\ref{fig:fit}.

Because they contain dynamics exclusive to experiments on ions in a Paul trap,
many of these models have in effect been tuned to that experiment.
They are thus not applicable to the other types of
experiments~\cite{petta_coherent_2005,cole_nature_2001,zrenner_nature_2002,
wang_macdonald_prb_2005,ramsay_prl_2010}. 
The 
models~\cite{romito_decoherence_2007,wang_macdonald_prb_2005,ramsay_prl_2010,
mogilevtsev_prl_2008}
used to explain the EID measured in
experiments with shallow donors and quantum dots have typically assumed
a feedback dynamics from, for example,  excitation induced phonons in 
substrates,
and these models have often been quantitatively successful.

When one invokes dynamics to explain a kinematic effect, of course, one risks
overly tuning models and misdiagnosing the underlying physics.
We have demonstrated that none of these dynamical mechanisms
is necessary to explain Excitation Induced Dephasing.
As shown in Appendix~\ref{app:rab},
an analogous model for distinguishable ensemble
members predicts no dependence of $\gamma$ on $\Omega$.
We therefore conclude that a dependence of $\gamma$ on $\Omega$,
or EID, may indeed be general, as suggested by experiment, and that it 
may be a kinematic effect
of the indistinguishability of separate, uncontrolled interactions
between quantum systems and their environment.

\section{Conclusion}
In addition to the typical reduced dynamics for open quantum
systems~\cite{petruccione_open_quantum_systems}, to understand
environmental interference one must faithfully represent the physics
involved in the preparation, evolution, and measurement of all members of the
experimental ensemble.
This means one must address the possibility that 
different subsets of the experimental ensemble may suffer 
environmental interference differently.

We have introduced a framework to treat multiple
quantum environments.
We have discussed a kinematic effect of indistinguishability that
affects only open quantum systems.
The indistinguishability of experimental ensemble members greatly complicates
the partitioning that our approach requires.
Based on the beam splitter model, we have made the simplifying 
hypothesis that environmental interference occurs in events
that are themselves indistinguishable.
We have not yet investigated alternative treatments, but
the kinematic nature of the effect would seem to preclude the possibility of
treating it dynamically through a single, ``effective'' environmental system
and a typical reduced dynamics.

To demonstrate our generalization for multiple quantum environments, 
within the framework we have described two simple systems.
We calculated correlation functions for photons scattered at a beam splitter,
with and without losses.
We created a decoherence model for systems undergoing Rabi oscillations.
In one case~\cite{meekhof_rabi_1996}, we have found unprecedented
quantitative agreement with measurements.
We have also discovered that the kinematic effect of indistinguishability
can explain the generally measured Excitation Induced Dephasing
that has previously required different dynamical explanations
for different experiments.
Full treatment of experiments will likely require more detailed models
addressing multiple sources of environmental interference, dressed states, 
reduced dynamics, etc.
But the qualitative and quantitative success
of our single, general model for different types of
Rabi oscillations experiments is very promising.

\appendix
\section{Including kinematics directly for photon scattering}
\label{app:pho}
Here we treat photon scattering as an open system and directly include
the kinematics of indistinguishability for a fixed value of $n$.
This corresponds to partitioning an ensemble of an arbitrary number of single
photons into an ensemble of an arbitrary number of $n$-photon objects.

Let $\rho_{A,n}$ represent the state of the $n$ actively prepared,
incident photons.
The density operator $\rho_{A,n}$ is defined in the tensor product space
$\Phi_A^{\otimes n}$, which is spanned by the
ket $\ket{i}_1\ket{i}_2\ldots\ket{i}_n\equiv\ket{i_1i_2\ldots i_n}$.
The operator $\rho_{A,n}$ therefore remains trivial even when $n$ photons
are incident.

The beam splitter passively prepares every member of the experimental ensemble
to be in a state represented by $\rho_{B,n}$, which is defined
in the tensor product space $\Phi_B^{\otimes n}$.
Again we will ignore entanglement, so
$\rho_{B,n}$ is a sum of projection operators into each of the
$2^n$ one-dimensional subspaces of $\Phi_B^{\otimes n}$.
Each projection operator is weighted by
$R^{n_R}T^{n_T}$, where $n_R$ and $n_T$ are the
numbers of reflected and transmitted photons, respectively, represented
by the associated one-dimensional subspace.

Consider, for example, the case where $n=3$.
One writes the properly symmetrized density operator, $\rho_{B,3}$, to be used
on the right hand side of~\eqref{eq:branching}, as
\begin{eqnarray} \label{eq:rb3}
\rho_{B,3}&=&R^3\ketbra{rrr}{rrr} \nonumber \\
& &+ R^2T\Big(\ketbra{rrt}{rrt}
+ \ketbra{rtr}{rtr} + \ketbra{trr}{trr}\Big) \nonumber \\
& & + RT^2\Big(\ketbra{rtt}{rtt} + \ketbra{trt}{trt} 
     + \ketbra{ttr}{ttr}\Big) \nonumber \\
& &+ T^3\ketbra{ttt}{ttt}.
\end{eqnarray}
If $R+T=1$, then $\mathrm{Tr}\rho_{B,n}=1$ for all $n$.

In the same basis, it is straightforward to construct operators
corresponding to the correlation functions calculated in
Section~\ref{sec:secquant}.
For example,
\begin{eqnarray} \label{eq:nrop}
\hat{n}_R & = & 3\,\ketbra{rrr}{rrr} \nonumber \\
 & & + 2\Big( \ketbra{rrt}{rrt}+\ketbra{rtr}{rtr}+\ketbra{trr}{trr}\Big)
 \nonumber \\
 & & + 1\Big( \ketbra{rtt}{rtt}+\ketbra{trt}{trt}+\ketbra{ttr}{ttr}\Big).
\end{eqnarray}
Proceeding in this fashion, one can reproduce~\eqref{eq:2quantcor}
for $n=3$.

\section{The Rabi oscillations model for distinguishable objects}
\label{app:rab}
Here we develop a model based on~\eqref{eq:cases} but for
distinguishable ensemble members and interference events.
Our scenario is as follows:
\begin{enumerate}
\item The members of the ensemble can again
suffer environmental interactions at the times
$n\Delta\tilde{t}$, where $n=1,2,3\ldots$
\item At every $n\Delta\tilde{t}$, there is some probability, $(1-\eta$),
for a member to suffer perturbation and thereby to be prepared passively.
\end{enumerate}
We have again used $\Delta\tilde{t}$ for the timescale of interference.
But now we have used the parameter, $\eta$, with $0\leq\eta\leq 1$,
to represent the susceptibility of ensemble members to 
environmental interference.
The parameter $\eta$ is the probability for any given
member of the ensemble not to suffer interference at one of the
times $n\Delta \tilde{t}$.
For a perfectly isolated system, $\eta=1$.

We will again assume that passive measurements are fair measurements,
and we will not assume any reduced dynamics.
We can then write a general formula for the Born probability
as more and more branches form:
\begin{equation}
\label{prob_written}
\mathcal{P}_{\ket{g}\bra{g}}\big(\rho(\tilde{t})\big) =
\begin{cases}
p_0(\tilde{t}) & 0\leq\tilde{t}< 1\,\Delta\tilde{t} \\
p_1(\tilde{t}) & 1\,\Delta\tilde{t}\leq\tilde{t}< 2\,\Delta\tilde{t} \\
\quad\vdots & \qquad\vdots \\
p_n(\tilde{t}) & n\Delta\tilde{t}\leq\tilde{t}< (n+1)\Delta\tilde{t} 
\end{cases}
\end{equation}
Here, every $p_n(\tilde{t})$ will have
the form of the sum in~\eqref{eq:sum_branch}, and $M_{branch}=n$.

If members are actively prepared to be in the excited state, and
because none will have suffered environmental interference before
$\tilde{t}=1\,\Delta\tilde{t}$, we have for the initial value
$p_0(\tilde{t})=\textrm{sin}^2(\Omega \tilde{t})$.
It is straightforward to find from the theory of Rabi oscillations
\begin{widetext}
\begin{equation}
\label{eq:first_rec}
p_1(\tilde{t})=\eta\,p_0(\tilde{t})+
(1-\eta)
\Big(p_0(1\Delta\tilde{t})\,
\textrm{cos}^2\big(\Omega\,(\tilde{t}-1\Delta\tilde{t})\big)
+ \big(1-p_0(1\Delta\tilde{t})\big)\,
\textrm{sin}^2\big(\Omega\,(\tilde{t}-1\Delta\tilde{t}) \big)
 \Big).
\end{equation}
For general $n$, we have
\begin{equation} \label{eq:general_rec}
p_n(\tilde{t})=\eta\,p_{n-1}(\tilde{t})+
(1-\eta)
\Big( p_{n-1}(n\Delta\tilde{t})\,
\textrm{cos}^2\big(\Omega\,(\tilde{t}-n\Delta\tilde{t})\big)
+\big(1-p_{n-1}(n\Delta\tilde{t})\big)\,
\textrm{sin}^2\big(\Omega\,(\tilde{t}-n\Delta\tilde{t})
\big) \Big).
\end{equation}
\end{widetext}
Note the reappearance of a recursive structure similar to
that of~\eqref{eq:two_events}.
Recursively partitioning the experimental ensemble
in this way~\eqref{eq:general_rec}, however, requires that one can
at any time label different ensemble members and know to which partition
they belong.
Because the recursion index, $n$, is also the index counting time in functions
of $n\Delta\tilde{t}$, there is no need to rescale the time parameter,
as was done in~\eqref{eq:trescale}.

\begin{figure*}[htp]
  \begin{center}
    \subfigure[With $\eta=0.99$, we fit $\gamma/\Omega=0.05$.]
    {\label{rec1}\includegraphics[width=0.45\textwidth]{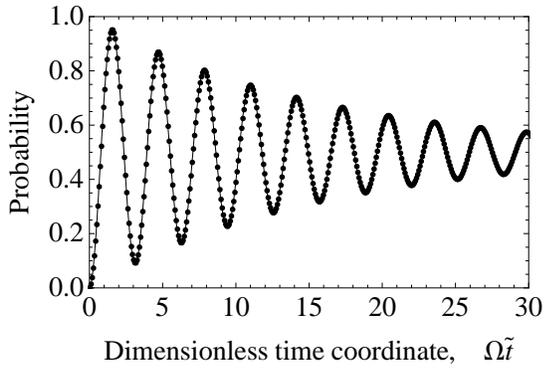}}\hfill
    \subfigure[With $\eta=0.997$, we fit $\gamma/\Omega=0.015$.]
    {\label{rec2}\includegraphics[width=0.45\textwidth]{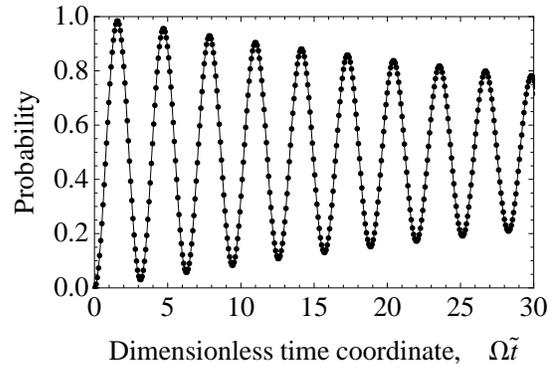}}\\
  \end{center}
  \caption{The Born probability,
           $\mathcal{P}_{\ket{g}\bra{g}}\big(\rho(\tilde{t})\big)$,
           when members of the ensemble are distinguishable.  For both plots,
           $\Omega\Delta\tilde{t}\approx 0.08$.
           The dots are the results of recursive calculations
           using \eqref{eq:general_rec} and \eqref{prob_written}.
           The solid lines are plots of the damped sinusoid
           \eqref{damped_rabi} that fits the experimental data.
           Recall that $\eta$ is the probability that members
           will not suffer a perturbation at the times $n\Delta\tilde{t}$.}
  \label{fig:recursive}
\end{figure*}

Figure~\ref{fig:recursive} shows the results of two sample calculations using
\eqref{prob_written} with \eqref{eq:general_rec}.
The dots are calculated from our model.
The solid lines are plots of the decaying sinusoid in \eqref{damped_rabi},
which fits the experimental measurements.
For both figures we have used $\Omega\Delta\tilde{t}\approx 0.08$.
The results in Figure~\ref{rec1} were calculated using $\eta=0.99$
and resulted in a fitted value for the damping factor of
$\gamma/\Omega=0.05$.
The results in Figure~\ref{rec2} were calculated using $\eta=0.997$
and resulted in a fitted value for the damping factor of
$\gamma/\Omega=0.015$.
Recall that $\eta=1$ for a perfectly isolated system.

This calculation for distinguishable members results in a fitted
damping factor, $\gamma$, that is independent of $\Omega$.

\bibliography{/home/pbryant/Documents/SMWork/QM/qm_bibliography}

\begin{thebibliography}{10}%
\makeatletter
\providecommand \@ifxundefined [1]{%
 \ifx #1\undefined \expandafter \@firstoftwo
 \else \expandafter \@secondoftwo
\fi
}%
\providecommand \@ifnum [1]{%
 \ifnum #1\expandafter \@firstoftwo
 \else \expandafter \@secondoftwo
\fi
}%
\providecommand \enquote [1]{``#1''}%
\providecommand \bibnamefont  [1]{#1}%
\providecommand \bibfnamefont [1]{#1}%
\providecommand \citenamefont [1]{#1}%
\providecommand\href[0]{\@sanitize\@href}%
\providecommand\@href[1]{\endgroup\@@startlink{#1}\endgroup\@@href}%
\providecommand\@@href[1]{#1\@@endlink}%
\providecommand \@sanitize [0]{\begingroup\catcode`\&12\catcode`\#12\relax}%
\@ifxundefined \pdfoutput {\@firstoftwo}{%
 \@ifnum{\z@=\pdfoutput}{\@firstoftwo}{\@secondoftwo}%
}{%
 \providecommand\@@startlink[1]{\leavevmode\special{html:<a href="#1">}}%
 \providecommand\@@endlink[0]{\special{html:</a>}}%
}{%
 \providecommand\@@startlink[1]{%
  \leavevmode
  \pdfstartlink
   attr{/Border[0 0 1 ]/H/I/C[0 1 1]}%
   user{/Subtype/Link/A<</Type/Action/S/URI/URI(#1)>>}%
  \relax
 }%
 \providecommand\@@endlink[0]{\pdfendlink}%
}%
\providecommand \url  [0]{\begingroup\@sanitize \@url }%
\providecommand \@url [1]{\endgroup\@href {#1}{\urlprefix}}%
\providecommand \urlprefix [0]{URL }%
\providecommand \Eprint[0]{\href }%
\@ifxundefined \urlstyle {%
  \providecommand \doi [1]{doi:\discretionary{}{}{}#1}%
}{%
  \providecommand \doi [0]{doi:\discretionary{}{}{}\begingroup
  \urlstyle{rm}\Url }%
}%
\providecommand \doibase [0]{http://dx.doi.org/}%
\providecommand \Doi[1]{\href{\doibase#1}}%
\providecommand \bibAnnote [3]{%
  \BibitemShut{#1}%
  \begin{quotation}\noindent
    \textsc{Key:}\ #2\\\textsc{Annotation:}\ #3%
  \end{quotation}%
}%
\providecommand \bibAnnoteFile [2]{%
  \IfFileExists{#2}{\bibAnnote {#1} {#2} {\input{#2}}}{}%
}%
\providecommand \typeout [0]{\immediate \write \m@ne }%
\providecommand \selectlanguage [0]{\@gobble}%
\providecommand \bibinfo [0]{\@secondoftwo}%
\providecommand \bibfield [0]{\@secondoftwo}%
\providecommand \translation [1]{[#1]}%
\providecommand \BibitemOpen[0]{}%
\providecommand \bibitemStop [0]{}%
\providecommand \bibitemNoStop [0]{.\EOS\space}%
\providecommand \EOS [0]{\spacefactor3000\relax}%
\providecommand \BibitemShut [1]{\csname bibitem#1\endcsname}%
\bibitem{dirac_qm_book}%
  \BibitemOpen
  \bibfield{author}{%
  \bibinfo {author} {\bibfnamefont{P.~A.~M.}\ \bibnamefont{{Dirac}}},\ }%
  \emph{\bibinfo {title} {The Principles of Quantum Mechanics}},\ \bibinfo
  {edition} {4th}\ ed.\ (\bibinfo {publisher} {Oxford University Press},\
  \bibinfo {year} {1958})%
  \bibAnnoteFile{NoStop}{dirac_qm_book}%
\bibitem{petruccione_open_quantum_systems}%
  \BibitemOpen
  \bibfield{author}{%
  \bibinfo {author} {\bibfnamefont{H.}~\bibnamefont{Breuer}}\ and\ \bibinfo
  {author} {\bibfnamefont{F.}~\bibnamefont{Petruccione}},\ }%
  \emph{\bibinfo {title} {The Theory of Open Quantum Systems}}\ (\bibinfo
  {publisher} {Oxford University Press},\ \bibinfo {year} {2002})%
  \bibAnnoteFile{NoStop}{petruccione_open_quantum_systems}%
\bibitem{meekhof_rabi_1996}%
  \BibitemOpen
  \bibfield{author}{%
  \bibinfo {author} {\bibfnamefont{D.~M.}\ \bibnamefont{Meekhof}}, \bibinfo
  {author} {\bibfnamefont{C.}~\bibnamefont{Monroe}}, \bibinfo {author}
  {\bibfnamefont{B.~E.}\ \bibnamefont{King}}, \bibinfo {author}
  {\bibfnamefont{W.~M.}\ \bibnamefont{Itano}},\ and\ \bibinfo {author}
  {\bibfnamefont{D.~J.}\ \bibnamefont{Wineland}},\ }%
  \bibfield{journal}{%
  \Doi{{10.1103/PhysRevLett.76.1796}}{\bibinfo {journal} {Phys. Rev. Lett.}}\
  }%
  \textbf{\bibinfo {volume} {76}},\ \bibinfo {pages} {1796} (\bibinfo {month}
  {Mar.}\ \bibinfo {year} {1996})%
  \bibAnnoteFile{NoStop}{meekhof_rabi_1996}%
\bibitem{nagourney_dehmelt_shelved_1986}%
  \BibitemOpen
  \bibfield{author}{%
  \bibinfo {author} {\bibfnamefont{W.}~\bibnamefont{Nagourney}}, \bibinfo
  {author} {\bibfnamefont{J.}~\bibnamefont{Sandberg}},\ and\ \bibinfo {author}
  {\bibfnamefont{H.}~\bibnamefont{Dehmelt}},\ }%
  \bibfield{journal}{%
  \bibinfo {journal} {Phys. Rev. Lett.}\ }%
  \textbf{\bibinfo {volume} {56}},\ \bibinfo {pages} {2797} (\bibinfo {month}
  {Jun.}\ \bibinfo {year} {1986})%
  \bibAnnoteFile{NoStop}{nagourney_dehmelt_shelved_1986}%
\bibitem{bergquist_qjumps_1986}%
  \BibitemOpen
  \bibfield{author}{%
  \bibinfo {author} {\bibfnamefont{J.~C.}\ \bibnamefont{Bergquist}}, \bibinfo
  {author} {\bibfnamefont{R.~G.}\ \bibnamefont{Hulet}}, \bibinfo {author}
  {\bibfnamefont{W.~M.}\ \bibnamefont{Itano}},\ and\ \bibinfo {author}
  {\bibfnamefont{D.~J.}\ \bibnamefont{Wineland}},\ }%
  \bibfield{journal}{%
  \bibinfo {journal} {Phys. Rev. Lett.}\ }%
  \textbf{\bibinfo {volume} {57}},\ \bibinfo {pages} {1699} (\bibinfo {month}
  {Oct}\ \bibinfo {year} {1986})%
  \bibAnnoteFile{NoStop}{bergquist_qjumps_1986}%
\bibitem{sauter_toschek_qjumps_1986}%
  \BibitemOpen
  \bibfield{author}{%
  \bibinfo {author} {\bibfnamefont{T.}~\bibnamefont{Sauter}}, \bibinfo {author}
  {\bibfnamefont{W.}~\bibnamefont{Neuhauser}}, \bibinfo {author}
  {\bibfnamefont{R.}~\bibnamefont{Blatt}},\ and\ \bibinfo {author}
  {\bibfnamefont{P.~E.}\ \bibnamefont{Toschek}},\ }%
  \bibfield{journal}{%
  \bibinfo {journal} {Phys. Rev. Lett.}\ }%
  \textbf{\bibinfo {volume} {57}},\ \bibinfo {pages} {1696} (\bibinfo {month}
  {Oct}\ \bibinfo {year} {1986})%
  \bibAnnoteFile{NoStop}{sauter_toschek_qjumps_1986}%
\bibitem{peik_qjumps_1994}%
  \BibitemOpen
  \bibfield{author}{%
  \bibinfo {author} {\bibfnamefont{E.}~\bibnamefont{Peik}}, \bibinfo {author}
  {\bibfnamefont{G.}~\bibnamefont{Hollemann}},\ and\ \bibinfo {author}
  {\bibfnamefont{H.}~\bibnamefont{Walther}},\ }%
  \bibfield{journal}{%
  \bibinfo {journal} {Phys. Rev. A}\ }%
  \textbf{\bibinfo {volume} {49}},\ \bibinfo {pages} {402} (\bibinfo {month}
  {Jan}\ \bibinfo {year} {1994})%
  \bibAnnoteFile{NoStop}{peik_qjumps_1994}%
\bibitem{brune_rabi_1996}%
  \BibitemOpen
  \bibfield{author}{%
  \bibinfo {author} {\bibfnamefont{M.}~\bibnamefont{Brune}}, \bibinfo {author}
  {\bibfnamefont{F.}~\bibnamefont{{Schmidt-Kaler}}}, \bibinfo {author}
  {\bibfnamefont{A.}~\bibnamefont{Maali}}, \bibinfo {author}
  {\bibfnamefont{J.}~\bibnamefont{Dreyer}}, \bibinfo {author}
  {\bibfnamefont{E.}~\bibnamefont{Hagley}}, \bibinfo {author}
  {\bibfnamefont{J.~M.}\ \bibnamefont{Raimond}},\ and\ \bibinfo {author}
  {\bibfnamefont{S.}~\bibnamefont{Haroche}},\ }%
  \bibfield{journal}{%
  \Doi{{10.1103/PhysRevLett.76.1800}}{\bibinfo {journal} {Phys. Rev. Lett.}}\
  }%
  \textbf{\bibinfo {volume} {76}},\ \bibinfo {pages} {1800} (\bibinfo {month}
  {Mar.}\ \bibinfo {year} {1996})%
  \bibAnnoteFile{NoStop}{brune_rabi_1996}%
\bibitem{wineland_experimental_1998}%
  \BibitemOpen
  \bibfield{author}{%
  \bibinfo {author} {\bibfnamefont{D.~J.}\ \bibnamefont{Wineland}}, \bibinfo
  {author} {\bibfnamefont{C.}~\bibnamefont{Monroe}}, \bibinfo {author}
  {\bibfnamefont{W.~M.}\ \bibnamefont{Itano}}, \bibinfo {author}
  {\bibfnamefont{D.}~\bibnamefont{Leibfried}}, \bibinfo {author}
  {\bibfnamefont{B.~E.}\ \bibnamefont{King}},\ and\ \bibinfo {author}
  {\bibfnamefont{D.~M.}\ \bibnamefont{Meekhof}},\ }%
  \bibfield{journal}{%
  \bibinfo {journal} {J. Res. Natl. Inst. Stand. Technol.}\ }%
  \textbf{\bibinfo {volume} {103}},\ \bibinfo {pages} {259} (\bibinfo {month}
  {Jun.}\ \bibinfo {year} {1998})%
  \bibAnnoteFile{NoStop}{wineland_experimental_1998}%
\bibitem{leggett_arrow_qmeasurement}%
  \BibitemOpen
  \bibfield{author}{%
  \bibinfo {author} {\bibfnamefont{A.}~\bibnamefont{Leggett}},\ }%
  in\ \emph{\bibinfo {booktitle} {Time's arrows today: recent physical and
  philosophical work on the direction of time}},\ \bibinfo {editor} {edited by\
  \bibinfo {editor} {\bibfnamefont{S.~F.}\ \bibnamefont{Savitt}}}\ (\bibinfo
  {publisher} {Cambridge University Press},\ \bibinfo {year} {1997})%
  \bibAnnoteFile{NoStop}{leggett_arrow_qmeasurement}%
\bibitem{stamatescu_collapse_2009}%
  \BibitemOpen
  \bibfield{author}{%
  \bibinfo {author} {\bibfnamefont{I.~O.}\ \bibnamefont{Stamatescu}},\ }%
  in\ \emph{\bibinfo {booktitle} {Compendium of Quantum Physics}},\ \bibinfo
  {editor} {edited by\ \bibinfo {editor}
  {\bibfnamefont{D.}~\bibnamefont{Greenberger}}, \bibinfo {editor}
  {\bibfnamefont{K.}~\bibnamefont{Hentschel}},\ and\ \bibinfo {editor}
  {\bibfnamefont{F.}~\bibnamefont{Weinert}}}\ (\bibinfo {publisher}
  {Springer},\ \bibinfo {year} {2009})\ pp.\ \bibinfo {pages} {813--822}%
  \bibAnnoteFile{NoStop}{stamatescu_collapse_2009}%
\bibitem{dicke_superrad_1954}%
  \BibitemOpen
  \bibfield{author}{%
  \bibinfo {author} {\bibfnamefont{R.~H.}\ \bibnamefont{Dicke}},\ }%
  \bibfield{journal}{%
  \bibinfo {journal} {Phys. Rev.}\ }%
  \textbf{\bibinfo {volume} {93}},\ \bibinfo {pages} {99} (\bibinfo {year}
  {1954})%
  \bibAnnoteFile{NoStop}{dicke_superrad_1954}%
\bibitem{lidar_decofree_2001}%
  \BibitemOpen
  \bibfield{author}{%
  \bibinfo {author} {\bibfnamefont{J.}~\bibnamefont{Kempe}}, \bibinfo {author}
  {\bibfnamefont{D.}~\bibnamefont{Bacon}}, \bibinfo {author}
  {\bibfnamefont{D.}~\bibnamefont{Lidar}},\ and\ \bibinfo {author}
  {\bibfnamefont{K.}~\bibnamefont{Whaley}},\ }%
  \bibfield{journal}{%
  \bibinfo {journal} {Phys. Rev. A}\ }%
  \textbf{\bibinfo {volume} {63}},\ \bibinfo {pages} {042307} (\bibinfo {month}
  {Mar.}\ \bibinfo {year} {2001})%
  \bibAnnoteFile{NoStop}{lidar_decofree_2001}%
\bibitem{buttiker_92}%
  \BibitemOpen
  \bibfield{author}{%
  \bibinfo {author} {\bibfnamefont{M.}~\bibnamefont{B\"uttiker}},\ }%
  \bibfield{journal}{%
  \bibinfo {journal} {Phys. Rev. B}\ }%
  \textbf{\bibinfo {volume} {46}},\ \bibinfo {pages} {12485} (\bibinfo {year}
  {1992})%
  \bibAnnoteFile{NoStop}{buttiker_92}%
\bibitem{loudon_91}%
  \BibitemOpen
  \bibfield{author}{%
  \bibinfo {author} {\bibfnamefont{R.}~\bibnamefont{Loudon}},\ }%
  in\ \emph{\bibinfo {booktitle} {Disorder in Condensed Matter Physics}},\
  \bibinfo {editor} {edited by\ \bibinfo {editor}
  {\bibfnamefont{J.}~\bibnamefont{Blackman}}\ and\ \bibinfo {editor}
  {\bibfnamefont{J.}~\bibnamefont{Tag\"ue\~na}}}\ (\bibinfo {publisher}
  {Clarendon Press},\ \bibinfo {address} {Oxford},\ \bibinfo {year} {1991})\
  p.\ \bibinfo {pages} {441}%
  \bibAnnoteFile{NoStop}{loudon_91}%
\bibitem{gerry_optics_book}%
  \BibitemOpen
  \bibfield{author}{%
  \bibinfo {author} {\bibfnamefont{C.~C.}\ \bibnamefont{Gerry}}\ and\ \bibinfo
  {author} {\bibfnamefont{P.~L.}\ \bibnamefont{Knight}},\ }%
  \emph{\bibinfo {title} {Introductory quantum optics}}\ (\bibinfo {publisher}
  {Cambridge University Press},\ \bibinfo {year} {2005})%
  \bibAnnoteFile{NoStop}{gerry_optics_book}%
\bibitem{dodd_atoms_1991}%
  \BibitemOpen
  \bibfield{author}{%
  \bibinfo {author} {\bibfnamefont{J.~N.}\ \bibnamefont{Dodd}},\ }%
  \emph{\bibinfo {title} {Atoms and Light: Interactions}}\ (\bibinfo
  {publisher} {Plenum Press},\ \bibinfo {address} {New York},\ \bibinfo {year}
  {1991})\ ISBN \bibinfo {isbn} {0306437414}%
  \bibAnnoteFile{NoStop}{dodd_atoms_1991}%
\bibitem{petta_coherent_2005}%
  \BibitemOpen
  \bibfield{author}{%
  \bibinfo {author} {\bibfnamefont{J.~R.}\ \bibnamefont{Petta}}, \bibinfo
  {author} {\bibfnamefont{A.~C.}\ \bibnamefont{Johnson}}, \bibinfo {author}
  {\bibfnamefont{J.~M.}\ \bibnamefont{Taylor}}, \bibinfo {author}
  {\bibfnamefont{E.~A.}\ \bibnamefont{Laird}}, \bibinfo {author}
  {\bibfnamefont{A.}~\bibnamefont{Yacoby}}, \bibinfo {author}
  {\bibfnamefont{M.~D.}\ \bibnamefont{Lukin}}, \bibinfo {author}
  {\bibfnamefont{C.~M.}\ \bibnamefont{Marcus}}, \bibinfo {author}
  {\bibfnamefont{M.~P.}\ \bibnamefont{Hanson}},\ and\ \bibinfo {author}
  {\bibfnamefont{A.~C.}\ \bibnamefont{Gossard}},\ }%
  \bibfield{journal}{%
  \Doi{10.1126/science.1116955}{\bibinfo {journal} {Science}}\ }%
  \textbf{\bibinfo {volume} {309}},\ \bibinfo {pages} {2180} (\bibinfo {month}
  {Sep.}\ \bibinfo {year} {2005})%
  \bibAnnoteFile{NoStop}{petta_coherent_2005}%
\bibitem{cole_nature_2001}%
  \BibitemOpen
  \bibfield{author}{%
  \bibinfo {author} {\bibfnamefont{B.~E.}\ \bibnamefont{Cole}}, \bibinfo
  {author} {\bibfnamefont{J.~B.}\ \bibnamefont{Williams}}, \bibinfo {author}
  {\bibfnamefont{B.~T.}\ \bibnamefont{King}}, \bibinfo {author}
  {\bibfnamefont{M.~S.}\ \bibnamefont{Sherwin}},\ and\ \bibinfo {author}
  {\bibfnamefont{C.~R.}\ \bibnamefont{Stanley}},\ }%
  \bibfield{journal}{%
  \bibinfo {journal} {Nature}\ }%
  \textbf{\bibinfo {volume} {410}},\ \bibinfo {pages} {60} (\bibinfo {month}
  {Mar.}\ \bibinfo {year} {2001})%
  \bibAnnoteFile{NoStop}{cole_nature_2001}%
\bibitem{zrenner_nature_2002}%
  \BibitemOpen
  \bibfield{author}{%
  \bibinfo {author} {\bibfnamefont{A.}~\bibnamefont{Zrenner}}, \bibinfo
  {author} {\bibfnamefont{E.}~\bibnamefont{Beham}}, \bibinfo {author}
  {\bibfnamefont{S.}~\bibnamefont{Stufler}}, \bibinfo {author}
  {\bibfnamefont{F.}~\bibnamefont{Findeis}}, \bibinfo {author}
  {\bibfnamefont{M.}~\bibnamefont{Bichler}},\ and\ \bibinfo {author}
  {\bibfnamefont{G.}~\bibnamefont{Abstreiter}},\ }%
  \bibfield{journal}{%
  \bibinfo {journal} {Nature}\ }%
  \textbf{\bibinfo {volume} {418}},\ \bibinfo {pages} {612} (\bibinfo {year}
  {2002})%
  \bibAnnoteFile{NoStop}{zrenner_nature_2002}%
\bibitem{wang_macdonald_prb_2005}%
  \BibitemOpen
  \bibfield{author}{%
  \bibinfo {author} {\bibfnamefont{Q.~Q.}\ \bibnamefont{Wang}}, \bibinfo
  {author} {\bibfnamefont{A.}~\bibnamefont{Muller}}, \bibinfo {author}
  {\bibfnamefont{P.}~\bibnamefont{Bianucci}}, \bibinfo {author}
  {\bibfnamefont{E.}~\bibnamefont{Rossi}}, \bibinfo {author}
  {\bibfnamefont{Q.~K.}\ \bibnamefont{Xue}}, \bibinfo {author}
  {\bibfnamefont{T.}~\bibnamefont{Takagahara}}, \bibinfo {author}
  {\bibfnamefont{C.}~\bibnamefont{Piermarocchi}}, \bibinfo {author}
  {\bibfnamefont{A.~H.}\ \bibnamefont{{MacDonald}}},\ and\ \bibinfo {author}
  {\bibfnamefont{C.~K.}\ \bibnamefont{Shih}},\ }%
  \bibfield{journal}{%
  \bibinfo {journal} {Phys. Rev. B}\ }%
  \textbf{\bibinfo {volume} {72}},\ \bibinfo {pages} {035306} (\bibinfo {month}
  {Jul.}\ \bibinfo {year} {2005})%
  \bibAnnoteFile{NoStop}{wang_macdonald_prb_2005}%
\bibitem{ramsay_prl_2010}%
  \BibitemOpen
  \bibfield{author}{%
  \bibinfo {author} {\bibfnamefont{A.~J.}\ \bibnamefont{Ramsay}}, \bibinfo
  {author} {\bibfnamefont{A.~V.}\ \bibnamefont{Gopal}}, \bibinfo {author}
  {\bibfnamefont{E.~M.}\ \bibnamefont{Gauger}}, \bibinfo {author}
  {\bibfnamefont{A.}~\bibnamefont{Nazir}}, \bibinfo {author}
  {\bibfnamefont{B.~W.}\ \bibnamefont{Lovett}}, \bibinfo {author}
  {\bibfnamefont{A.~M.}\ \bibnamefont{Fox}},\ and\ \bibinfo {author}
  {\bibfnamefont{M.~S.}\ \bibnamefont{Skolnick}},\ }%
  \bibfield{journal}{%
  \bibinfo {journal} {Phys. Rev. Lett.}\ }%
  \textbf{\bibinfo {volume} {104}},\ \bibinfo {pages} {017402} (\bibinfo {year}
  {2010})%
  \bibAnnoteFile{NoStop}{ramsay_prl_2010}%
\bibitem{murao_decoherence_1998}%
  \BibitemOpen
  \bibfield{author}{%
  \bibinfo {author} {\bibfnamefont{M.}~\bibnamefont{Murao}}\ and\ \bibinfo
  {author} {\bibfnamefont{P.~L.}\ \bibnamefont{Knight}},\ }%
  \bibfield{journal}{%
  \bibinfo {journal} {Phys. Rev. A}\ }%
  \textbf{\bibinfo {volume} {58}},\ \bibinfo {pages} {663} (\bibinfo {year}
  {1998})%
  \bibAnnoteFile{NoStop}{murao_decoherence_1998}%
\bibitem{bonifacio_pra_2000}%
  \BibitemOpen
  \bibfield{author}{%
  \bibinfo {author} {\bibfnamefont{R.}~\bibnamefont{Bonifacio}}, \bibinfo
  {author} {\bibfnamefont{S.}~\bibnamefont{Olivares}}, \bibinfo {author}
  {\bibfnamefont{P.}~\bibnamefont{Tombesi}},\ and\ \bibinfo {author}
  {\bibfnamefont{D.}~\bibnamefont{Vitali}},\ }%
  \bibfield{journal}{%
  \bibinfo {journal} {Phys. Rev. A}\ }%
  \textbf{\bibinfo {volume} {61}},\ \bibinfo {pages} {053802} (\bibinfo {month}
  {Apr.}\ \bibinfo {year} {2000})%
  \bibAnnoteFile{NoStop}{bonifacio_pra_2000}%
\bibitem{difidio_damped_2000}%
  \BibitemOpen
  \bibfield{author}{%
  \bibinfo {author} {\bibfnamefont{C.}~\bibnamefont{Di~Fidio}}\ and\ \bibinfo
  {author} {\bibfnamefont{W.}~\bibnamefont{Vogel}},\ }%
  \bibfield{journal}{%
  \bibinfo {journal} {Phys. Rev. A}\ }%
  \textbf{\bibinfo {volume} {62}},\ \bibinfo {pages} {031802(R)} (\bibinfo
  {year} {2000})%
  \bibAnnoteFile{NoStop}{difidio_damped_2000}%
\bibitem{schneider_decoherence_1998}%
  \BibitemOpen
  \bibfield{author}{%
  \bibinfo {author} {\bibfnamefont{S.}~\bibnamefont{Schneider}}\ and\ \bibinfo
  {author} {\bibfnamefont{G.~J.}\ \bibnamefont{Milburn}},\ }%
  \bibfield{journal}{%
  \bibinfo {journal} {Phys. Rev. A}\ }%
  \textbf{\bibinfo {volume} {57}},\ \bibinfo {pages} {3748} (\bibinfo {year}
  {1998})%
  \bibAnnoteFile{NoStop}{schneider_decoherence_1998}%
\bibitem{budini_localization_2002}%
  \BibitemOpen
  \bibfield{author}{%
  \bibinfo {author} {\bibfnamefont{A.~A.}\ \bibnamefont{Budini}}, \bibinfo
  {author} {\bibfnamefont{R.~L.}\ \bibnamefont{de~Matos~Filho}},\ and\ \bibinfo
  {author} {\bibfnamefont{N.}~\bibnamefont{Zagury}},\ }%
  \bibfield{journal}{%
  \bibinfo {journal} {Phys. Rev. A}\ }%
  \textbf{\bibinfo {volume} {65}},\ \bibinfo {pages} {041402(R)} (\bibinfo
  {month} {Apr.}\ \bibinfo {year} {2002})%
  \bibAnnoteFile{NoStop}{budini_localization_2002}%
\bibitem{budini_dissipation_2003}%
  \BibitemOpen
  \bibfield{author}{%
  \bibinfo {author} {\bibfnamefont{A.~A.}\ \bibnamefont{Budini}}, \bibinfo
  {author} {\bibfnamefont{R.~L.}\ \bibnamefont{de~Matos~Filho}},\ and\ \bibinfo
  {author} {\bibfnamefont{N.}~\bibnamefont{Zagury}},\ }%
  \bibfield{journal}{%
  \bibinfo {journal} {Phys. Rev. A}\ }%
  \textbf{\bibinfo {volume} {67}},\ \bibinfo {pages} {033815} (\bibinfo {month}
  {Mar.}\ \bibinfo {year} {2003})%
  \bibAnnoteFile{NoStop}{budini_dissipation_2003}%
\bibitem{serra_decoherence_2001}%
  \BibitemOpen
  \bibfield{author}{%
  \bibinfo {author} {\bibfnamefont{R.~M.}\ \bibnamefont{Serra}}, \bibinfo
  {author} {\bibfnamefont{N.~G.}\ \bibnamefont{de~Almeida}}, \bibinfo {author}
  {\bibfnamefont{W.~B.}\ \bibnamefont{da~Costa}},\ and\ \bibinfo {author}
  {\bibfnamefont{M.~H.~Y.}\ \bibnamefont{Moussa}},\ }%
  \bibfield{journal}{%
  \bibinfo {journal} {Phys. Rev. A}\ }%
  \textbf{\bibinfo {volume} {64}},\ \bibinfo {pages} {033419} (\bibinfo {year}
  {2001})%
  \bibAnnoteFile{NoStop}{serra_decoherence_2001}%
\bibitem{romito_decoherence_2007}%
  \BibitemOpen
  \bibfield{author}{%
  \bibinfo {author} {\bibfnamefont{A.}~\bibnamefont{Romito}}\ and\ \bibinfo
  {author} {\bibfnamefont{Y.}~\bibnamefont{Gefen}},\ }%
  \bibfield{journal}{%
  \Doi{{10.1103/PhysRevB.76.195318}}{\bibinfo {journal} {Phys. Rev. B}}\ }%
  \textbf{\bibinfo {volume} {76}},\ \bibinfo {pages} {195318} (\bibinfo {month}
  {Nov.}\ \bibinfo {year} {2007})%
  \bibAnnoteFile{NoStop}{romito_decoherence_2007}%
\bibitem{mogilevtsev_prl_2008}%
  \BibitemOpen
  \bibfield{author}{%
  \bibinfo {author} {\bibfnamefont{D.}~\bibnamefont{Mogilevtsev}}, \bibinfo
  {author} {\bibfnamefont{A.~P.}\ \bibnamefont{Nisovtsev}}, \bibinfo {author}
  {\bibfnamefont{S.}~\bibnamefont{Kilin}}, \bibinfo {author}
  {\bibfnamefont{S.~B.}\ \bibnamefont{Cavalcanti}}, \bibinfo {author}
  {\bibfnamefont{H.~S.}\ \bibnamefont{Brandi}},\ and\ \bibinfo {author}
  {\bibfnamefont{L.~E.}\ \bibnamefont{Oliveira}},\ }%
  \bibfield{journal}{%
  \bibinfo {journal} {Phys. Rev. Lett.}\ }%
  \textbf{\bibinfo {volume} {100}},\ \bibinfo {pages} {017401} (\bibinfo {year}
  {2008})%
  \bibAnnoteFile{NoStop}{mogilevtsev_prl_2008}%
\end{thebibliography}%

\end{document}